\documentclass[AMA,STIX1COL]{WileyNJD-v2}

\articletype{RESEARCH ARTICLE}
\received{9 August 2018}
\revised{5 November 2018}
\accepted{10 November 2018}
\doi{10.1002/nme.5998}
\usepackage{Paper}

\begin{document}

\title{Transported snapshot model order reduction approach for parametric,
steady-state fluid flows containing parameter dependent shocks}

\author[1]{Nirmal J. Nair*}

\author[1]{Maciej Balajewicz}

\authormark{NAIR AND BALAJEWICZ}

\address{\orgdiv{Department of Aerospace Engineering}, \orgname{University of Illinois at Urbana-Champaign}, \orgaddress{\city{Urbana}, \state{Illinois}, \country{USA}}}

\corres{*Nirmal J. Nair, \orgdiv{Department of Aerospace Engineering}, \orgname{University of Illinois at Urbana-Champaign}, \orgaddress{\city{Urbana}, \state{Illinois}, \country{USA}}. 
\\ \email{njn2@illinois.edu}}

\fundingInfo{Air Force Office of Scientific Research, Grant/Award Number: FA9550-17-1-0203}

\abstract[Summary]{A new model order reduction approach is proposed for parametric steady-state
nonlinear fluid flows characterized by shocks and discontinuities whose spatial
locations and orientations are strongly parameter dependent. In this method,
solutions in the predictive regime are approximated using a linear
superposition of parameter dependent basis. The sought after parametric
reduced-basis are obtained by transporting the snapshots in a spatially and parametrically dependent transport field. Key to the
proposed approach is the observation that the transport fields are
typically smooth and continuous, despite the solution themselves not being so.
As a result, the transport fields can be accurately expressed using a
low-order polynomial expansion. Similar to traditional projection-based model order reduction
approaches, the proposed method is formulated mathematically as a residual
minimization problem for the generalized coordinates. The proposed approach is also integrated with well-known hyper-reduction strategies to obtain significant computational speed-ups.
The method is successfully applied to the reduction of a parametric 1-D flow in a
converging-diverging nozzle, a parametric 2-D supersonic flow over a forward
facing step and a parametric 2-D jet diffusion flame in a combustor.}


\jnlcitation{\cname{%
\author{Nair NJ},
\author{Balajewicz M.}} 
\ctitle{Transported snapshot model order reduction approach for parametric, steady-state fluid flows containing parameter dependent shocks}. \cjournal{Int J Numer Methods Eng}. \cyear{2018;1-29}.
\url{https://doi.org/10.1002/nme.5998}}

\maketitle

 
\section{Introduction}

Computational models of high-dimensional systems arise in a rich variety of
engineering and scientific contexts. Computational Fluid Dynamics (CFD) for
example has become an indispensable tool for many engineering applications
across a wide range of industries.  Unfortunately, high-fidelity CFD simulations
are often so computationally prohibitive that they cannot be used as often as
needed or used only in special circumstances rather than routinely.
Consequently, the impact of CFD on parametric and time-critical applications
such as design, optimization, and control has not yet been fully
realized. Model Order Reduction (MOR) is a serious contender for bridging this
gap. 

Most existing MOR approaches are based on projection. In projection-based MOR,
the state variables are approximated in a low-dimensional subspace. Over the
years, a number of approaches for calculating a reduced-order basis (ROB) have
been developed, e.g., proper orthogonal decomposition
(POD)~\cite{galletti2004low,holmes2012turbulence}, dynamic mode decomposition
(DMD)~\cite{rowley2009spectral,schmid2010dynamic}, balanced POD
(BPOD)~\cite{rowley2005model,willcox2002balanced}, balanced
truncation~\cite{gugercin2004survey,moore1981principal} and the reduced-basis
method~\cite{rozza2011reduced,veroy2005certified}. 

For a reduced-order model (ROM) to be truly useful, it must be capable of generating accurate
predictions for parameter values that are different from those sampled for the
purpose of constructing a ROB. Generating ROBs and ROMs that are robust to
parameter variations is an active area of research~\cite{benner2015survey}. The
choice of parameter sample points is critical to any method used to generate
the basis. For problems with smaller number of parameters, a simple approach
like random sampling using the Latin hypercube method is often
sufficient~\cite{iman2014latin}.  For problems with large number of
parameters, more sophisticated sampling methods are usually required.  In the
standard greedy sampling
approach~\cite{paul2015adaptive,rozza2008reduced,bui2008parametric,bui2008model},
the sample points are chosen one-by-one in an adaptive manner. At every
iteration, the goal is to find the parameter value for which the error between
the ROM and the full order model (FOM) is largest. The FOM is sampled at this point and the new
information is used to generate a new reduced-basis.

Another critical issue involves the choice between a global or local basis.
Although global basis have been shown to perform adequately for many
applications~\cite{rewienski2006model,lieu2006reduced,amsallem2010toward,
carlberg2011efficient}, particularly challenging problems often necessitate the
use of multiple local reduced
basis~\cite{yano2014space,dahmen2015best}. In these
cases, several local basis are constructed and linked to particular regions in
the parameter or state
space~\cite{washabaugh2012nonlinear,amsallem2012nonlinear,zahr2013construction}.
The price of this additional flexibility is the switching algorithm required
to switch between the local basis. Finally, to improve the ROM performance in
predictive regimes, it is also possible to interpolate the basis or the ROM
matrices~\cite{amsallem2011online,amsallem2009method,bui2003proper}.

Achieving parametric robustness is particularly challenging when the sought
after solutions contain sharp gradients, discontinuities or shocks. These
situations arise in a wide range of important engineering applications,
for example, high-speed fluid
flows~\cite{franz2014interpolation,hummel2016reduced}, multi-phase flows with
evolving material interfaces~\cite{balajewicz2014reduction}, computational
finance~\cite{balajewicz2016reduced} and structural contact problem with
evolving contact regions~\cite{washabaugh2016use}. 

Over the years, a large variety of discontinuity-aware MOR techniques have been
developed. In the first class of such methods, the problem of modeling
discontinuities is avoided entirely by exploiting symmetry and transport
reversal properties of certain hyperbolic
PDEs~\cite{rim2018transport,mowlavi2018model,rowley2000reconstruction}.  Other
methods involve decomposition into global and advection modes governed by
optimal mass transfer~\cite{iollo2014advection} or more direct modeling of
discontinuities using basis splitting~\cite{carlberg2015adaptive} or snapshot
transformation~\cite{welper2017interpolation,reiss2018shifted,cagniart2016model}. For unsteady solutions with
shocks, accurate low-rank solutions can be obtained using a Lagrangian framework
~\cite{mojgani2017lagrangian} where both position and state of Lagrangian
particle variables are approximated by their respective ROBs.  Other methods
decompose the solution into a variable separable form consisting of an evolution
term to capture moving shocks and a diffusion term to capture the changing
shapes~\cite{ohlberger2013nonlinear}. Finally, other methods avoid the problem of
modeling discontinuities by domain decomposition where direct numerical
simulation or reconstruction methods are used in regions containing the
discontinuities
~\cite{lucia2001reduced,constantine2012reduced,cagniart2017model}. 

In this manuscript we outline our new proposed approach for parametric model
reduction of solutions containing moving shocks and discontinuities. In our
proposed approach, solutions in the predictive regime are approximated using parameter dependent basis. Snapshots are transported in a spatially and parametrically dependent transport field to yield transported snapshots. These transported snapshots form the parameter dependent basis for the proposed MOR. Key to our proposed approach is the
observation that the transport fields are smooth and thus, can be
themselves approximated using a low-order expansion. Our method may be
interpreted as a data-driven generalization of previous works
of Rowley and Marsden~\cite{rowley2000reconstruction} and Iollo and
Lombardi~\cite{iollo2014advection}.

The remainder of the paper is organized as follows.  In \S~\ref{sec:Problem
statement}, the problem of interest and the traditional projection-based model
order reduction approach is recapitulated.  The proposed methodology for
approximating unsampled solutions via transported snapshots for one-dimensional parameter variational problems is detailed  in
\S~\ref{TSMOR_1D} and its extension to multi-dimensions is detailed in \S~\ref{TSMOR}. The details for numerical implementation of our new approach in the discrete framework are provided in \S~\ref{Implementation}.
In \S~\ref{HR}, the proposed approach is integrated with a hyper-reduction algorithm. In \S~\ref{Comparison}, the novelty of the proposed approach is compared with similar existing methods in literature. In
\S~\ref{Numericalexperiments}, the performance of the proposed method is
evaluated on several simple but representative fluid flow models. Limitations of the proposed approach are discussed in \S~\ref{Limitations}. Finally, in
\S~\ref{Conclusions}, conclusions are offered and prospects for future work
are summarized.


\section{Problem statement}
\label{sec:Problem statement}

\subsection{Full order model}

We consider full order models consisting of
hyperbolic or convection-dominated parabolic PDEs such as the Euler or
high-Reynolds-number Navier-Stokes equations:
\begin{equation}
  \frac{\partial \bm{u}}{\partial t}+\bm{f}(\bm{u},\bm{x};\bm{\mu})=\bm{0}
  \label{unsteady1}
\end{equation}
where the state variable $\bm{u}=\bm{u}(\bm{x},t;\bm{\mu})$ depends on space $\bm{x}\in \Omega$, $\Omega$ being the flow domain, time $t \in [0,t_{\text{max}}]$ and a vector of $N_d$ parameters $\bm{\mu} \in \mathcal{D} \subset \mathbb{R}^{N_d}$
(where $\mathcal{D}$ is a bounded domain). $\bm{f}$ is the nonlinear function that  contains the convective and diffusive fluxes.

The steady state equation for this system can
be written as:
\begin{equation}
  \bm{R}\left(\bm{u};\bm{\mu}\right) \coloneqq \bm{f}(\bm{u},\bm{x};\bm{\mu})=\bm{0}
  \label{steady1}
\end{equation}
where $\bm{R}(\bm{u};\bm{\mu})$ is the steady state residual.  To obtain steady state
solutions, Eq.~\eqref{steady1} can be discretized in space by a standard finite difference/volume/element method. The resulting set of equations can be solved directly by an iterative method or a
time-stepping method can be used to advance the semi-discretized form of unsteady Eq.
\eqref{unsteady1} to a steady state.


\subsection{Traditional nonlinear model reduction}
\label{Traditional}
In traditional projection-based MOR, the state variable $\bm{u}(\bm{x};\bm{\mu})$ is approximated in a global low-dimensional trial subspace as follows
\begin{equation}
  \bm{u}(\bm{x};\bm{\mu}) 
    \approx \bm{u}_r = \sum_{n=1}^{N_k} \bm{u}_n(\bm{x}) a_n(\bm{\mu})
  \label{MOR1}
\end{equation}
where $\bm{u}_n$ is the basis of this subspace, $N_k$ is the number of basis, and
$\bm{a}(\bm{\mu}) \in \mathbb{R}^{N_k}$ denotes the generalized coordinates in
this basis. Substituting the approximation \eqref{MOR1} into the residual equation~\eqref{steady1} yields
$\bm{R}\left(\bm{u}_r;\bm{\mu}\right)=\bm{0}$.
Consequently, the generalized coordinates are chosen to minimize the
residual of the Galerkin expansion.
\begin{equation}
  \left(\bm{v}_n, \bm{R}(\bm{u}_r;\bm{\mu})\right)_\Omega
\label{MOR3}
\end{equation}
where $\bm{v}_n$ is the basis of the test subspace and the inner product is defined as
\begin{equation}
\left(\bm{u},\bm{v} \right)_\Omega:=\int_\Omega \bm{u}.\bm{v} d\bm{x}
\label{inner}
\end{equation}

If the test basis $\bm{v}_n \neq \bm{u}_n$, then the projection in Eq.~\eqref{MOR3} is called as Petrov-Galerkin projection with the specific case $\bm{v}_n =\bm{J}\bm{u}_n$ known as least-squares Petrov-Galerkin (LSPG) projection, where $\bm{J}=\partial \bm{R}/\partial \bm{a}$ denotes the Jacobian of the residual. For nonlinear, non-self adjoining problems such as those represented in
this case by the set of ODEs, this approach is more robust than a Galerkin
projection, where $\bm{v}_n = \bm{u_n}$. 

Solving the minimization problem in equation~\eqref{MOR3} requires the evaluation of the residual of the
governing equations of the high-dimensional state variables. The complexity of
this computation scales with the size of the FOM. Therefore, while MOR approximates the FOM in a low-dimensional subspace, part of the computational cost still scales with the size of the FOM.
For general nonlinear systems, an additional level of approximation -- sometimes
called ``hyper-reduction'' such as the discrete empirical interpolation method (DEIM)~\cite{chaturantabut2010nonlinear}, Gauss-Newton with approximated tensors (GNAT) method~\cite{carlberg2011gnat}, energy-conserving sampling and weighting (ECSW) method~\cite{farhat2014dimensional}
-- is therefore required.

\subsection{Drawbacks}
\label{Drawbacks}

In data-driven projection-based MOR, the reduced-order bases are usually
constructed offline by collecting solution snapshots $\bm{u}(\bm{x};\bm{\mu}_s)$ of  
problem~\eqref{steady1} for different instances $\bm{\mu}_s$ for $s=1,\cdots,N_s$, of the parameter
vector $\bm{\mu}$.  
The reduced-order bases are then
formed by selecting a small subset of the snapshots or via compression using,
for example, POD.

For a ROM to be useful however, it must be capable of providing solutions at
parameters $\bm{\mu}^*$ not sampled during the offline basis construction
phase, $\bm{\mu}^* \neq \bm{\mu}_s$. Although parameter robustness is an active
area of research, it is particularly challenging when the sought after solutions
contain discontinuities or sharp gradients whose spatial orientations are
strongly parameter dependent. 

\begin{figure}
\centering
\begin{subfigure}[t]{0.45\textwidth}
\centering
\includegraphics[scale=1]{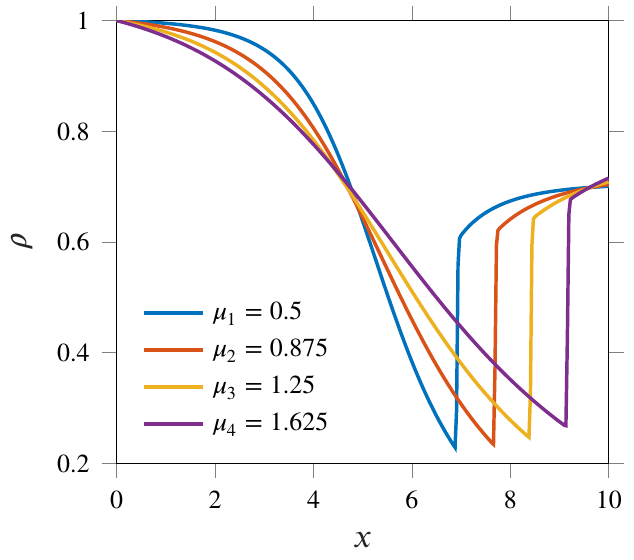}
\caption{Steady density solution $\rho$ for various throat area parameters ${\mu}_s$}
\label{4sol1}
\end{subfigure}\hspace{1cm}
\begin{subfigure}[t]{0.45\textwidth}
\centering
\includegraphics[scale=1]{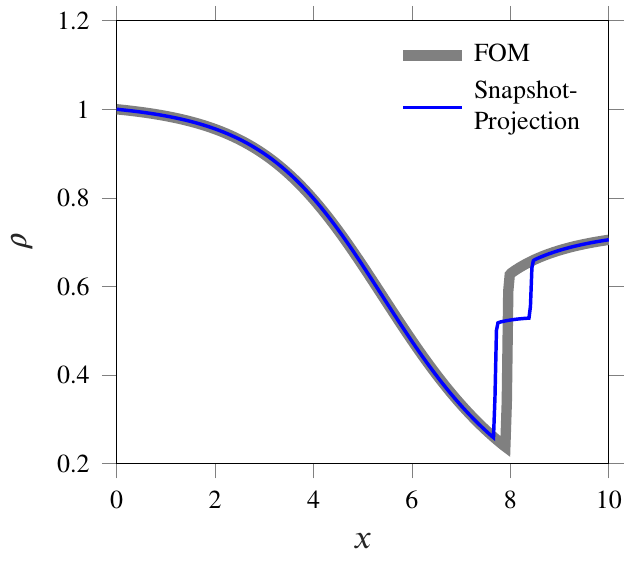}
\caption{Comparison of optimally constructed solution by Snapshot-Projection with FOM at $\mu^*=1.0$}
\label{nozzle_predict_trad}
\end{subfigure}
\caption{Steady state solutions at $\mu_s$ and optimally constructed solution at $\mu^*$}
\end{figure}

To illustrate, consider the simple problem of a quasi 1-D supersonic flow in a
converging-diverging nozzle governed by 1-D Euler equations. Area profile, $A(x)$, of the
nozzle is parabolic with equal inlet and outlet area, $A(0)=A(L)$, and the
throat is located at $L/2$, where $L$ is the length of the nozzle. For this
problem, the throat area $\mu=A(L/2)$ is the varying parameter of interest. Refer to
\S~\ref{problemdescription1} for details of this problem. Four snapshots at parameters $\mu_s=[0.5, 0.875, 1.25, 1.625]$
are generated and the corresponding steady density solutions are shown in
Fig.~\ref{4sol1}.

Optimal construction of a new solution at an unsampled parameter $\mu^*$ is
given by a superposition of the sampled snapshots:
\begin{equation}
\bm{u}_r(\bm{x};\mu^*)=\sum_{s=1}^{N_s}\bm{u}(\bm{x};\mu_s)a_n(\mu^*)
\end{equation}
where the coordinates, $\bm{a}(\mu^*)$, are
obtained by projecting the FOM solution onto the snapshots. The FOM and optimally constructed solutions at $\mu^*=1.0$ ($\mu_2<\mu^*<\mu_3$) are plotted in Fig.~\ref{nozzle_predict_trad}. It can be observed that the optimally constructed solution is dominated by staircase shock type error. Moreover, the optimal coordinates are $\bm{a}(\mu^*)=[-0.006, 0.333, 0.688, -0.016]$ where the coordinates corresponding to the first and fourth snapshots, ($a_1$ and $a_4$, respectively), are significantly lower than the coordinates corresponding to the second and third snapshots, ($a_2$ and $a_3$, respectively). This implies that the optimal construction is significantly dominated by local nearby snapshots.

This toy problem demonstrates that: (a) optimal constructions
of solutions characterized by parameter dependent shocks and discontinuities are
typically local and sparse in the sense that only two bases were required for the construction, and (b) the optimal construction provides a ``staircase'' approximation of the true solution. In summary, neither global nor local reduced-basis can be expected to yield
efficient approximations in the predictive regime of solutions characterized by
shocks, discontinuities, and sharp gradients whose physical locations and
orientations are parameter dependent.

\section{Transported Snapshot Model Order Reduction (TSMOR)}
\label{TSMOR_1D}

In this manuscript, we introduce and summarize our new MOR approach for parametric
and steady nonlinear fluid flows characterized by moving shocks, discontinuities
and sharp gradients. In this section, the proposed method is initially developed and detailed for one-dimensional parameter variational problems, while the extension of this method for multi-dimensional parameter variations are later detailed in \S~\ref{TSMOR}.

Our proposed approach is motivated by the observation that for many problems of
interest, such as the motivational problem illustrated in Fig.~\ref{4sol1},
snapshots local in the parameter space are transported in physical space. More
precisely, if $\bm{u}(\bm{x};\mu_{j})$ and $\bm{u}(\bm{x};\mu_{j+k})$ are the
sampled solutions of the FOM at parameters $\mu_j$ and $\mu_{j+k}$,
respectively, then there exists continuous and smooth transport field
$\bm{c}_j(\bm{x},\Delta \mu)$ such that
\begin{equation}
  \bm{u}(\bm{x};\mu_{j+k}) \approx \bm{u}(\bm{x} + \bm{c}_j(\bm{x},\Delta \mu);\mu_j)
\label{snapshot_pair}
\end{equation}
where $\Delta \mu = \mu_{j+k} - \mu_{j}$ is the parameter variation
between the original snapshot $\bm{u}(\bm{x},\mu_j)$ and the target snapshot
$\bm{u}(\bm{x},\mu_{j+k})$. 

Approximating the solution at an {\it unsampled} parameter  $\mu^* \neq \mu_s$,
for $s = 1,\ldots,N_s$ proceeds similarly to the traditional projection-based
MOR. More specifically, we assume that the solution can be approximated as a
linear superposition of parameter dependent basis functions
\begin{equation}
  \bm{u}(\bm{x};\mu^*) \approx \bm{u}_r = \sum_{n=1}^{N_k} \bm{u}_n(\bm{x};\mu^*) a_n(\mu^*)
  \label{TSMOR1}
\end{equation}
where the reduced-basis functions $\bm{u}_n$ correspond to
the {\it transported} local snapshots
\begin{equation}
  \bm{u}_n = \bm{u}(\bm{x} + \bm{c}_{k_n}(\bm{x},\Delta \mu);\mu_{k_n}) \quad \text{for} \quad n=1,\ldots,N_k
\label{shifted_snapshots}
\end{equation}
where the transport field $\bm{c}_{k_n}(\bm{x};\Delta \mu)$ is evaluated at the unsampled parameter variation, $\Delta
\mu = \mu^* - \mu_{k_n}$ and the snapshots, $\bm{u}(\bm{x},\mu_{k_n})$ for
$n=1,\cdots,N_k$, are a subset of the solution snapshots computed offline, ${\mu}_{k_n} \in {\mu}_{s}$.

Finally, the generalized coordinates, $\bm{a}(\mu^*)$, are
identified by minimizing the residual of the Petrov-Galerkin expansion. 
\begin{equation}
  \left(\bm{v}_n, \bm{R}(\bm{u}_r;\mu^*)\right)_\Omega
\label{MOR5}
\end{equation}
where $(\cdot,\cdot)_\Omega$ is the standard inner product as defined in Eq.~\eqref{inner}.

In summary, a new solution at an unsampled parameter using our proposed approach is given by a linear superposition of transported snapshots. This approach can be decomposed into the standard 
offline-online strategy where the transport fields are identified offline while the residual of the Petrov-Galerkin projection is minimized in the online stage.

\emph{Remark 1.} The transport fields in Eq.~\eqref{shifted_snapshots} are identified offline via a training procedure. Clearly, any direct
identification procedure for these fields would be intractable for large-scale
systems and, very likely, yield an over-determined and ill-conditioned system.
These computational issues can be avoided by adding smoothness constraints to the transport fields. Details of this procedure are
outlined in \S~\ref{Offline1}.

\emph{Remark 2.} For the sake of brevity, we have summarized the approach for 
one-dimensional parameter variations. However, the extension of this method to
multi-dimensional parameter variations are straightforward and the details of
this extension are provided in \S~\ref{TSMOR}.

\emph{Remark 3.} In practice, FOMs are typically derived in a semi-discrete form by discretizing a system of PDEs in space. Hence, usually, the solution
snapshots are available only as vectors and not as functions.
Consequently, it is not possible to simply {\it evaluate} the transported snapshots
for arbitrary transport field $\bm{c}_{k_n}(\bm{x},\Delta \mu)$ as shown in Eq.~\eqref{shifted_snapshots}. For discrete
models, this step must be performed using a numerical interpolation procedure.
Computational details of this procedure are outlined in \S~\ref{Implementation}. 

\emph{Remark 4.} Since the reduced-bases are parameter dependent, for
every new prediction at an unsampled parameter, an entire set of new reduced-bases must be
generated. However, the computational costs of generating these bases can be
expected to be proportional to the size of the FOM. Furthermore, the computational cost associated with the computation of the nonlinear residual function also scales with the size of the FOM. This expensive cost of residual evaluation can be mitigated by precomputation of certain terms that contain polynomial nonlinearities. However, such a precomputation procedure may not be viable for other classes of nonlinear functions. Hence, to gain significant amounts of computational speed-up,
a hyper-reduction strategy must be utilized. Details of this procedure are
outlined in \S~\ref{HR}.


\subsection{Offline stage}
\label{Offline1}

Given the set of snapshots at sampled parameters ${\mu}_s$, the methodology for identifying the transport fields $\bm{c}_s(\bm{x};\Delta \mu)$, for
$s=1,\cdots,N_s$, where $\Delta \mu=\mu-\mu_s$, for each snapshot is
explained in this section.

As mentioned in Remark 1 of \S~\ref{TSMOR_1D}, to make the identification procedure for the transport field tractable for large systems, smoothness constraints are required. More precisely, since the transport fields are assumed to be smooth in space, they can be approximated using a low order polynomial expansion in space using spatial basis functions $\bm{f}_p(\bm{x})$. The basis functions $\bm{f}_p(\bm{x})$ are selected \emph{a priori} such as Chebyshev polynomials or Fourier modes. Then, similar to polynomial fitting, the transport field is expressed as a polynomial expansion in parameter space as:
\begin{equation}
\bm{c}_s(\bm{x};\Delta \mu)=  \sum_{p=1}^{N_p} \sum_{q=1}^{N_q} c_m^{s} \bm{f}_p(\bm{x})g_q(\Delta \mu) \quad \text{for} \quad m=1,\ldots,N_p N_q
  \label{transport_polynomial3}
\end{equation}
where $c_m^{s}$ are the coefficients of the expansion, $N_p$ and $N_q$ are the number of functions $\bm{f}_p(\bm{x})$ and $g_q(\Delta \mu)$, respectively, and $g_q(\Delta \mu)$ are the monomials given by:
\begin{equation} 
g_q({\Delta \mu})= (\Delta \mu)^{q} \quad \text{for} \quad q=1,\ldots,N_q
\label{distance1}
\end{equation}
where $\Delta \mu=\mu-\mu_s$.
The choice of the basis functions $\bm{f}_p(\bm{x})$ is problem-dependent, hence it is detailed in \S~\ref{Numericalexperiments} with the help of test problems.

The least-squares fitting procedure to identify the coefficients $\bm{c}^{s}:= c_m^{s}$ is now explained. Firstly, a subset of $N_v$ solution snapshots $\bm{u}(\bm{x};{\mu}_{v_n})$ for $n=1,\cdots,N_v$, where ${\mu}_{v_n} \in {\mu}_{s}$, is chosen. Here, the parameters ${\mu}_{v_n}$ are called as the training parameters for the parameter ${\mu}_{s}$ and the corresponding snapshots are called training snapshots. The procedure for selecting $N_v$ training snapshots from $N_s$ is described in \S~\ref{Training_snapshots1}. The coefficients $\bm{c}^{s}$ are evaluated by minimizing the least-squares training error between the transported snapshot $\bm{u}(\bm{x}+\bm{c}_s(\bm{x};\Delta \mu);{\mu}_s)$ and training snapshots $\bm{u}(\bm{x};{\mu}_{v_n})$:
\begin{equation}
\underset{\bm{c}^{s}}{\text{min}}
   \sum_{n=1}^{N_v} \left( \bm{\epsilon}_n, \bm{\epsilon}_n \right)_{\Omega} 
\label{MOR6}
\end{equation}
where $\bm{\epsilon}_n(\bm{x})=\bm{u}(\bm{x}+\bm{c}_s(\bm{x};\Delta \mu);{\mu}_s)-\bm{u}(\bm{x};{\mu}_{v_n})$; $\Delta \mu=\mu_{v_n}-\mu_s$ and $(\cdot,\cdot)_\Omega$ is the standard inner product as defined in Eq.~\eqref{inner}.
The optimization problem \eqref{MOR6} is solved $N_s$ times for the coefficients
$\bm{c}^{s}$ for $s=1,\cdots,N_s$ snapshots. 

In summary, the identification of the transport field in Eq.~\eqref{transport_polynomial3} is posed as a least-squares fitting problem for the coefficients of the polynomial fitting function.

\subsubsection{Choice of training snapshots}
\label{Training_snapshots1}

According to Eq.~\eqref{snapshot_pair}, a snapshot at $\mu_{j+k}$ can be approximated by transporting its neighboring snapshot at $\mu_j$. The accuracy of this approximation tends to decrease as the absolute value, $\left|k\right|$, increases. Hence, $N_v$ training snapshots are chosen which correspond to the $N_v$ neighboring snapshots to $\bm{u}(\bm{x};{\mu}_{s})$ in the parameter space.

\begin{figure}[h]
\centering
\includegraphics[scale=0.25]{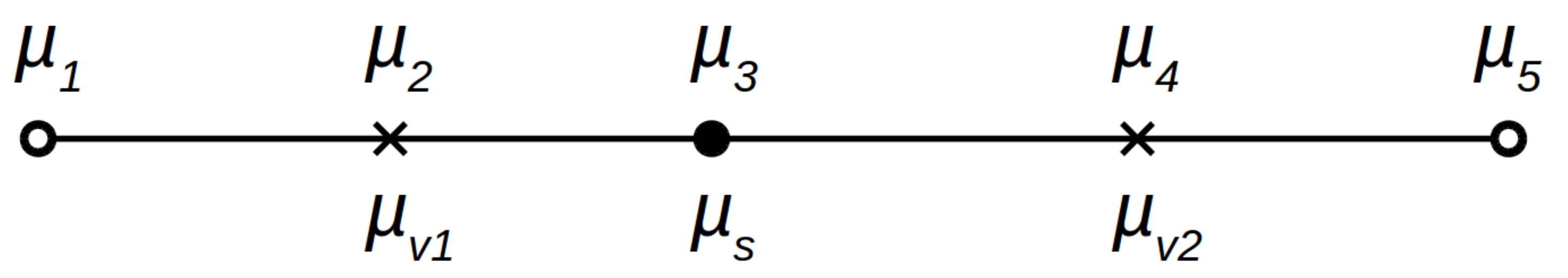}
\caption{Schematic for choosing training snapshots}
\label{par1D_1}
\end{figure}

For instance, consider the sampled parameters $\mu_1<\ldots<\mu_5$ in Fig.~\ref{par1D_1}. For the evaluation of the transport field for the snapshot at $\mu_3$, we have $\mu_s=\mu_3$. The snapshot at $\mu_3$ has $N_v=2$ nearest neighboring snapshots at $\mu_2$ and $\mu_4$ which form the training snapshots $\mu_{v_1}$ and $\mu_{v_2}$, respectively. It is noted that the offline training procedure does not restrict itself to $N_v=2$ and allows for more neighboring snapshots to be incorporated. Finally, a biased stencil can be used at the boundary of the parameter space for the selection of the training snapshots.
%


\subsection{Online stage}
In this stage, parameter dependent reduced-bases $\bm{u}_n(\bm{x};{\mu}^*)$ at an unsampled parameter ${\mu}^*$ ($\neq{\mu}_s$) are constructed and generalized coordinates $\bm{a}({\mu}^*)$ are identified.

For a parameter ${\mu}^*$ in the predictive regime, local reduced-bases $\bm{u}_n$ are evaluated by transporting local snapshots $\bm{u}(\bm{x};{\mu}_{k_n})$ for
$n=1,\cdots,N_k$, as shown in Eq.~\eqref{shifted_snapshots}, where ${\mu}_{k_n} \in {\mu}_{s}$ is a subset of snapshot solutions. The procedure for selecting the subset of snapshots $\bm{u}(\bm{x};{\mu}_{k_n})$ is described in \S~\ref{Basis}. For the predictive regime, the transport field $\bm{c}_{k_n}(\bm{x};\Delta \mu)$ in Eq.~\eqref{transport_polynomial3}, identified in the offline stage, is evaluated at the unsampled parameter, $\Delta
\mu = \mu^* - \mu_{k_n}$.

Finally, the generalized coordinates are obtained by solving the minimization problem~\eqref{MOR8}. The initial guesses for the generalized coordinates at the unsampled parameter $\bm{a}({\mu^*})^{(0)}$ are obtained by a linear combination of the generalized coordinates at the known snapshots, $\bm{a}({\mu}_{k_n})$. The generalized coordinates at the known snapshots are given by $\bm{a}({\mu}_{k_n})=\bm{e}_{k_n}$, where $\bm{e}_{k_n} \in \mathbb{R}^{N_k}$ is the $k_n$th canonical unit vector. The weights of the linear combination are obtained such that they are inversely proportional to $\left| {\Delta \mu} \right|$ and the sum of all weights equals one.

\subsubsection{Choice of basis} 
\label{Basis}

In traditional projection-based MOR applied to smooth elliptic problems, ROM
performance is expected to improve by increasing the number of bases $N_k$.
Unfortunately, this is usually not the case when the solutions of interest are
characterized by strong shocks and discontinuities. For
example, Abgrall et al.~\cite{abgrall2016robust} demonstrated that optimal constructions of
such solutions are typically  sparse in the generalized coordinates and local in
the parameter space. This property was also demonstrated in the nozzle problem considered in \S~\ref{Drawbacks}.
Therefore, in this work, we use only a small number of local bases corresponding to $N_k$ nearest neighboring snapshots
selected from the set of $N_s$ snapshots.

\begin{figure}[h]
\centering
\includegraphics[scale=0.25]{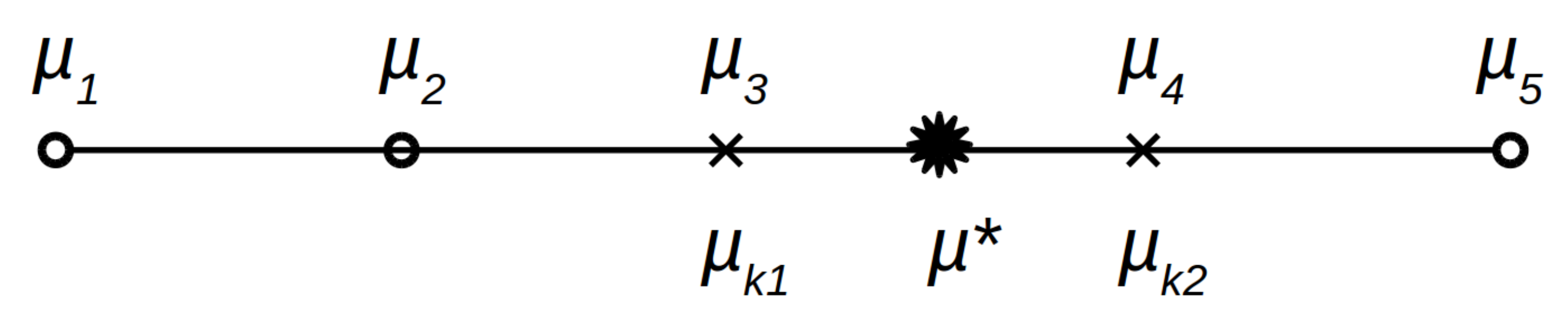}
\caption{Schematic for choosing the snapshots for basis construction}
\label{par1D_2}
\end{figure}

For instance, consider the same sampled parameters $\mu_1<\ldots<\mu_5$ in Fig.~\ref{par1D_2}. For the prediction at a new unsampled parameter $\mu^*$, $N_k=2$ nearest neighboring snapshots are given by $\mu_3$ and $\mu_4$ which form the snapshots $\mu_{k_1}$ and $\mu_{k_2}$, respectively, required for constructing the transported snapshots or local bases. It is noted that the proposed TSMOR approach does not restrict itself to $N_k=2$ bases and allows for more snapshots to be incorporated. Finally, for $\mu^*$ lying outside of the sampled parameter space (i.e. $\mu^*<\mu_1$ or $\mu^*>\mu_5$), a biased stencil can be used for the selection of the neighboring snapshots.
%


\section{Extension of TSMOR to multi-dimensional parameter variations}
\label{TSMOR}

In this section, the proposed TSMOR approach is detailed for multi-dimensional parameter variations where the snapshots are sampled on a Cartesian grid in parameter space, $\bm{\mu} \in \mathcal{D} \subset \mathbb{R}^{N_d}$.
Without loss of generality, the parameter vector $\bm{\mu}\in \mathbb{R}^{N_d}$ is normalized to fit in a hypercube such that every element of the parameter vector satisfies: $-1 \leq \mu_i \leq 1$.

TSMOR for multi-dimensional case proceeds exactly in the same manner as that for the one-dimensional case but with the difference that the one-dimensional parameter $\mu$ and parameter variation $\Delta \mu$ are replaced by the corresponding vectors $\bm{\mu}$ and $\bm{\Delta \mu}$, respectively. Consequently, solutions at an unsampled
parameter $\bm{\mu}^* \neq \bm{\mu}_s$ are approximated as a
linear superposition of parameter dependent basis functions
\begin{equation}
  \bm{u}(\bm{x};\bm{\mu}^*) \approx \bm{u}_r = \sum_{n=1}^{N_k} \bm{u}_n(\bm{x};\bm{\mu}^*) a_n(\bm{\mu}^*)
  \label{TSMOR2}
\end{equation}
where the reduced-basis functions, $\bm{u}_n$, correspond to
the {\it transported} local snapshots
\begin{equation}
  \bm{u}_n = \bm{u}(\bm{x} + \bm{c}_{k_n}(\bm{x},\bm{\Delta \mu});\bm{\mu}_{k_n}) \quad \text{for} \quad n=1,\ldots,N_k
\label{shifted_snapshots2}
\end{equation}
where the transport field $\bm{c}_{k_n}(\bm{x};\bm{\Delta \mu})$ is evaluated at the unsampled parameter variation, $\bm{\Delta \mu}=\bm{\mu}^*-\bm{\mu}_{k_n}$ and the snapshots, $\bm{u}(\bm{x},\mu_{k_n})$ for
$n=1,\cdots,N_k$, are a subset of the solution snapshots computed offline, $\bm{\mu}_{k_n} \in \bm{\mu}_{s}$. 

Finally, the generalized coordinates $\bm{a}(\bm{\mu}^*)$ are
identified by minimizing the residual of the Petrov-Galerkin expansion. 
\begin{equation}
  \left(\bm{v}_n, \bm{R}(\bm{u}_r;\bm{\mu}^*)\right)_\Omega
\label{MOR8}
\end{equation}
where $(\cdot,\cdot)_\Omega$ is the standard inner product as defined in Eq.~\eqref{inner}. 

The proposed TSMOR approach for the multi-dimensional case is also decomposed into the standard 
offline-online strategy where the transport fields are identified offline while the residual of the Petrov-Galerkin projection
is minimized in the online stage.

\subsection{Offline stage}
\label{Offline2}

Given the set of snapshots at sampled parameters $\bm{\mu}_s$, the methodology for identifying the transport fields $\bm{c}_s(\bm{x};\bm{\Delta \mu})$, for
$s=1,\cdots,N_s$, where $\bm{\Delta \mu}=\bm{\mu}-\bm{\mu}_{s}$, for each snapshot is
explained in this section.

Similar to TSMOR for one-dimensional parameter variations, the convection field is expressed as a low order expansion in space using spatial basis functions $\bm{f}_p(\bm{x})$. The basis functions $\bm{f}_p(\bm{x})$ are selected a priori such as Chebyshev polynomials
or Fourier modes. For the multi-dimensional case, the dependence of the transport field on parameter variation is modeled as a \emph{multi-variate} polynomial expansion in parameter space:
\begin{equation}
\bm{c}_s(\bm{x};\bm{\Delta \mu})=  \sum_{p=1}^{N_p} \sum_{q=1}^{N_q} c_m^{s} \bm{f}_p(\bm{x})g_q(\bm{\Delta \mu}) \quad \text{for} \quad m=1,\ldots,N_p N_q
  \label{transport_polynomial5}
\end{equation}
where $c_m^{s}$ are the coefficients of the expansion, $N_p$ and $N_q$ are the number of functions $\bm{f}_p(\bm{x})$ and $g_q(\bm{\Delta \mu})$, respectively, and $g_q(\bm{\Delta \mu})$ is given by:
\begin{equation} 
g_q(\bm{\Delta \mu})= \prod_{j=1}^{N_d}(\Delta \mu_j)^{h_{q,j}} \quad \text{for } h_{q,j} \in \mathbb{Z} \text{ and } q = 1,\ldots,N_q
\label{distance2}
\end{equation}
where ${\Delta \mu}_j$ for $j=1,\ldots,N_d$ are the vectorial
elements of $\bm{\Delta \mu}$. Few terms in Eq.~\eqref{transport_polynomial5} are expanded using Eq.~\eqref{distance2} for clarity as:
\begin{equation}
  \bm{c}_s(\bm{x};\bm{\Delta \mu}) = \sum_{p=1}^{N_p} c_m^{s} f_p(\bm{x})\Delta \mu_1  +  c_{m+1}^{s} f_p(\bm{x}) \Delta \mu_2  +  \ldots  + c_r^{s} f_p(\bm{x}) \Delta \mu_1^2  +  c_{r+1}^{s} f_p(\bm{x}) \Delta \mu_1 \mu_2 +  \ldots 
  \label{transport_polynomial4}
\end{equation}
Since the choice of $\bm{f}_p(\bm{x})$ is problem-dependent, it is detailed in \S~\ref{Numericalexperiments} with the help of test problems.

The least-squares fitting procedure to identify the coefficients $\bm{c}^{s}:= c_m^{s}$ is similar to the one-dimensional case. Firstly, a subset of solution snapshots $\bm{u}(\bm{x};\bm{\mu}_{v_n})$ for $n=1,\cdots,N_v$, where $\bm{\mu}_{v_n} \in \bm{\mu}_{s}$, is chosen. The procedure for selecting these $N_v$ training snapshots, $\bm{u}(\bm{x};\bm{\mu}_{v_n})$, from $N_s$ is described in \S~\ref{Training_snapshots2}. The coefficients $\bm{c}^{s}$ are evaluated by minimizing the least-squares training error between the transported snapshot $\bm{u}(\bm{x}+\bm{c}_s(\bm{x};\bm{\Delta \mu});\bm{\mu}_s)$ and training snapshots $\bm{u}(\bm{x};\bm{\mu}_{v_n})$:
\begin{equation}
  \underset{\bm{c}^{s}}{\text{min}}
   \sum_{n=1}^{N_v} \left( \bm{\epsilon}_n, \bm{\epsilon}_n \right)_{\Omega} \\
\label{MOR9}
\end{equation}
where $\bm{\epsilon}_n(\bm{x})=\bm{u}(\bm{x}+\bm{c}_s(\bm{x};\bm{\Delta \mu});\bm{\mu}_s)-\bm{u}(\bm{x};\bm{\mu}_{v_n})$; $\bm{\Delta \mu}=\bm{\mu}_{v_n}-\bm{\mu}_{s}$ and $(\cdot,\cdot)_\Omega$ is the standard inner product as defined in Eq.~\eqref{inner}.
The optimization problem \eqref{MOR9} is solved $N_s$ times for the coefficients
$\bm{c}^{s}$ for $s=1,\cdots,N_s$ snapshots. 

In summary, the transport field in Eq.~\eqref{transport_polynomial5} is identified by solving a least-squares fitting problem for the coefficients of the polynomial fitting function.

\subsubsection{Choice of training snapshots}
\label{Training_snapshots2}

Similar to one-dimensional parameter variations, $N_v$ training snapshots are chosen which correspond to the $N_v$ nearest neighboring snapshots to $\bm{u}(\bm{x};\bm{\mu}_{s})$. 

\begin{figure}[h]
\centering
\includegraphics[scale=0.26]{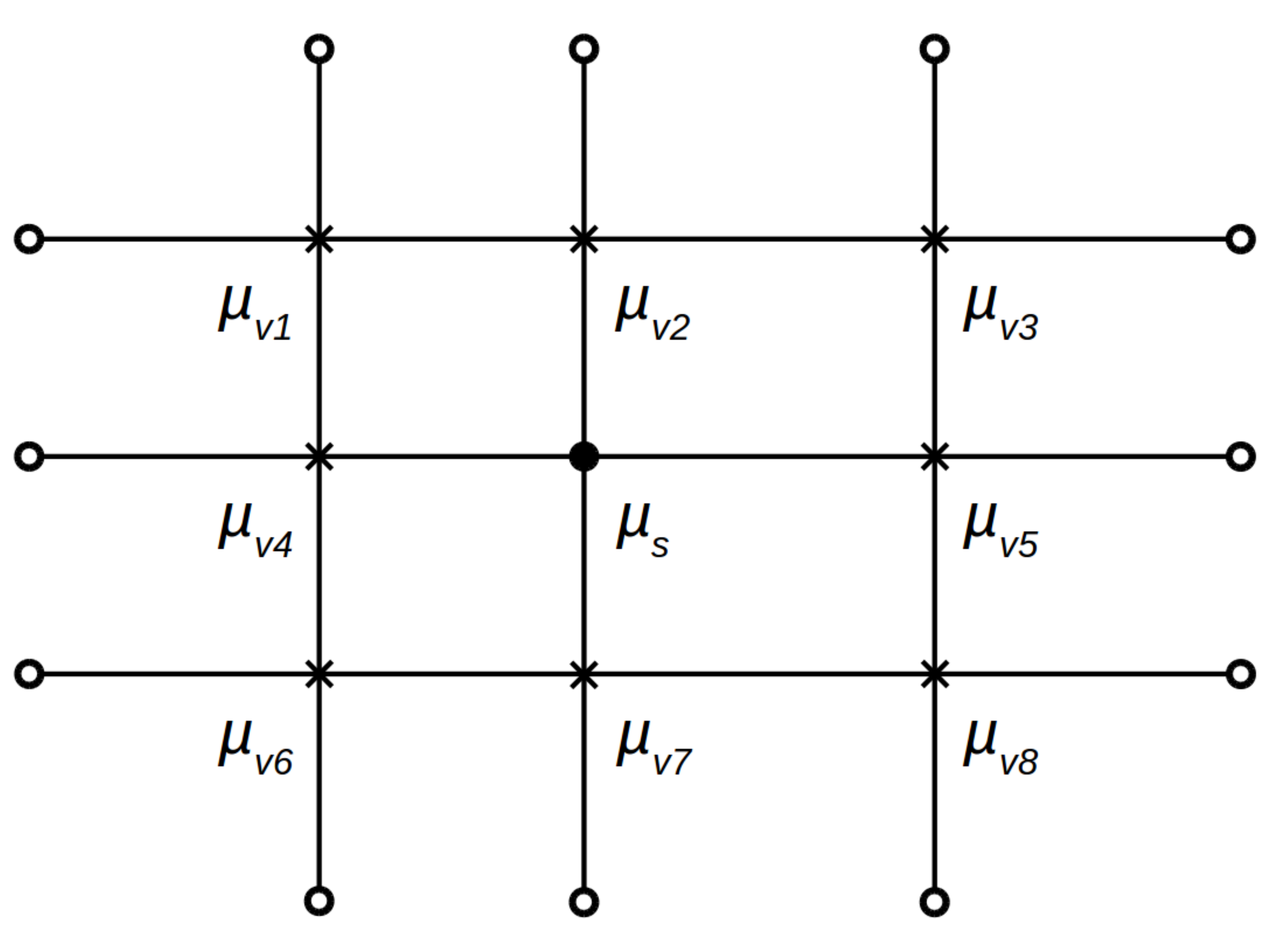}
\caption{Schematic for choosing training snapshots for 2-D parametrical case}
\label{par2D_1}
\end{figure}

For instance, consider the sampled parameters in a two-dimensional parameter space denoted by `$\times$', `$\circ$' and `$\bullet$' as shown in Fig.~\ref{par2D_1}. For the evaluation of the transport field for the snapshot at $\bm{\mu}_s$, $N_v=8$ nearest neighboring snapshots denoted by `$\times$' are chosen which form the training snapshots $\bm{\mu}_{v_1}$ through $\bm{\mu}_{v_8}$. The three-dimensional analogue would use $N_v=26$ neighboring snapshots for the offline training procedure and extends similarly for higher dimensions. Finally, a biased stencil can be used at the boundary of the parameter space for the selection of the training snapshots.
%


\subsection{Online stage}
In this stage, parameter dependent reduced-bases $\bm{u}_n(\bm{x};\bm{\mu}^*)$ at an unsampled parameter $\bm{\mu}^* \neq \bm{\mu}_s$ are constructed and generalized coordinates $\bm{a}(\bm{\mu}^*)$ are identified.

For a parameter $\bm{\mu}^*$ in the predictive regime, local reduced-bases $\bm{u}_n$ are evaluated by transporting local snapshots $\bm{u}(\bm{x};\bm{\mu}_{k_n})$ for
$n=1,\cdots,N_k$, as shown in Eq.~\eqref{shifted_snapshots2}, where $\bm{\mu}_{k_n} \in \bm{\mu}_{s}$ is a subset of snapshot solutions. The procedure for selecting the snapshots $\bm{u}(\bm{x};{\mu}_{k_n})$ is described in \S~\ref{Basis2}. For the predictive regime, the transport field $\bm{c}_{k_n}(\bm{x};\bm{\Delta \mu})$ in Eq.~\eqref{transport_polynomial5}, identified in the offline stage, is evaluated at the unsampled parameter, $\bm{\Delta \mu} = \bm{\mu}^* - \bm{\mu}_{k_n} $.

Finally, the generalized coordinates are obtained by solving the minimization problem~\eqref{MOR8}. The initial guesses for the generalized coordinates at the unsampled parameter $\bm{a}(\bm{\mu^*})^{(0)}$ are obtained by a linear combination of the generalized coordinates at the known snapshots, $\bm{a}(\bm{\mu}_{k_n})$. The generalized coordinates at the known snapshots are given by $\bm{a}(\bm{\mu}_{k_n})=\bm{e}_{k_n}$, where $\bm{e}_{k_n} \in \mathbb{R}^{N_k}$ is the $k_n$th canonical unit vector. The weights of the linear combination are obtained such that they are inversely proportional to $\left\| \bm{\Delta \mu} \right\|_2$ and the sum of all weights equals one.
\subsubsection{Choice of basis} 
\label{Basis2}

Similar to the one-dimensional case, we use only a small number of local bases corresponding to $N_k$ nearest neighboring snapshots
selected from the set of $N_s$ snapshots. 

\begin{figure}[h]
\centering
\includegraphics[scale=0.26]{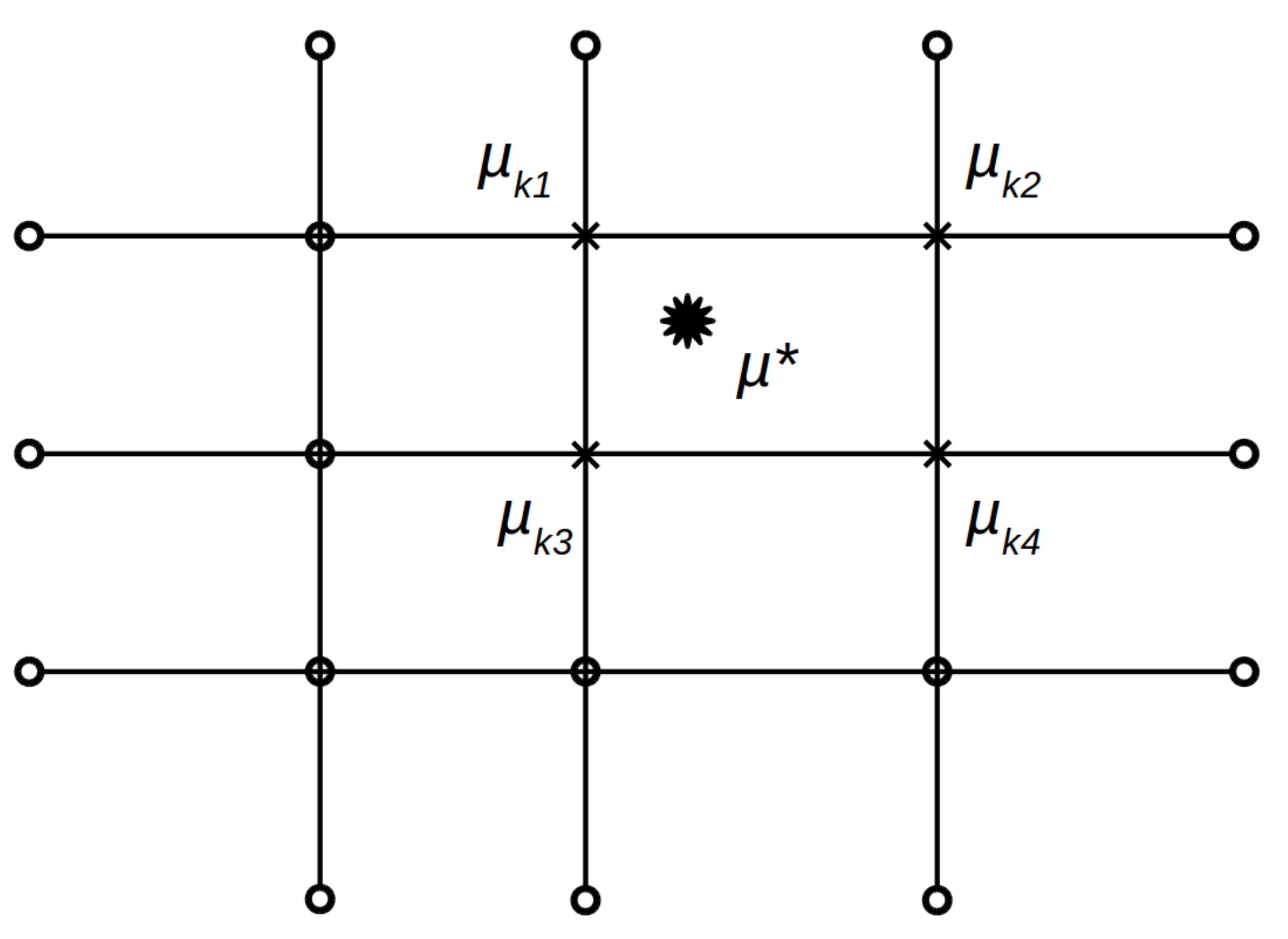}
\caption{Schematic for choosing the snapshots for basis construction for 2-D parametrical case}
\label{par2D_2}
\end{figure}

For instance, consider the same sampled parameters in a two-dimensional parameter space denoted by `$\times$' and `$\circ$' as shown in Fig.~\ref{par2D_2}. For the prediction at a new unsampled parameter $\bm{\mu}^*$, $N_k=4$ nearest neighboring snapshots denoted by `$\times$' are chosen which form the snapshots $\bm{\mu}_{k_1}$ through $\bm{\mu}_{k_4}$ required for constructing the transported snapshots or local bases. It is noted that the proposed TSMOR approach does not restrict itself to $N_k=4$ basis and allows for more snapshots to be incorporated for this two-dimensional case. The three-dimensional analogue would use $N_v=8$ neighboring snapshots for basis construction and extends similarly for higher dimensions. Finally, for $\mu^*$ lying outside of the sampled parameter space, a biased stencil can be used for the selection of the neighboring snapshots.
%


\section{Implementation of TSMOR in discrete framework}
\label{Implementation}

As mentioned in remark 3 of \S~\ref{TSMOR_1D}, FOM solutions are typically available as vectors of state variables specified at various spatial locations. Thus, this section discusses the implementation strategy for applying the proposed TSMOR approach in the discrete framework.

Let $\bm{w}(\bm{\mu}_s) \in \mathbb{R}^{N_w}$ denote the discrete FOM solution of the snapshot solution $\bm{u}(\bm{x};\bm{\mu}_s)$. In the continuous form, the transported snapshot $\bm{u}(\bm{x}+\bm{c}_s(\bm{x};\bm{\Delta \mu});\bm{\mu}_s)$, where $\bm{\Delta \mu}=\bm{\mu}-\bm{\mu}_s$, is directly evaluated by computing the solution at the new spatial locations $\bm{x}+\bm{c}_s(\bm{x};\bm{\Delta \mu})$. However, in the discrete form, an interpolation step is required to evaluate the transported snapshots. More specifically, if $\bm{w}(\bm{\mu}_{s})$ is the vector of state variables on a
computational grid with Cartesian coordinates $\bm{x}_i \in \mathbb{R}^{N_w}$
for $i =1,2,3$, then the transported
solution snapshot is evaluated by interpolating $\bm{w}(\bm{\mu}_{s})$ from $\bm{x}_i+\bm{c}_s(\bm{x}_i;\bm{\Delta \mu})$ to the original grid $\bm{x}_i$. In the discrete form, $\bm{c}_{s}(\bm{x}_i;\bm{\Delta \mu})\in \mathbb{R}^{N_w}$ denotes the transport field in Eq.~\eqref{transport_polynomial5} evaluated at $\bm{x}_i$ grid points and the transported snapshot from $\bm{\mu}_s$ to $\bm{\mu}$ is denoted by $\bm{w}'(\bm{\mu}_s,\bm{\mu}) \in \mathbb{R}^{N_w}$.

Consequently, the solution at the new parameter using the TSMOR approach is
given by
\begin{equation}
  \bm{w}(\bm{\mu}^*) 
    \approx \bm{w}_r(\bm{\mu}^*)
    =       \bm{U}(\bm{\mu}^*)\bm{a}(\bm{\mu}^*)
  \label{MOR44}
\end{equation}
where $\bm{U}(\bm{\mu}^*) \in \mathbb{R}^{N_w\times N_k}$ is a matrix whose
columns contain the \emph{transported} solution snapshots
$\bm{w}'(\bm{\mu}_{k_n},\bm{\mu}^*)$ of the corresponding snapshots
$\bm{w}(\bm{\mu}_{k_n})$. More precisely, $\bm{U}(\bm{\mu}^*)_{:,n} := \bm{w}'(\bm{\mu}_{k_n},\bm{\mu}^*)$, where the subscript $(:,n)$ denotes $n$th column of a matrix. The generalized
coordinates $\bm{a}(\bm{\mu^*})$ are
chosen to minimize the residual at the new parameter $\bm{\mu}^*$:
\begin{equation}
  \underset{\bm{a}}{\text{min}}
    \lVert \bm{R}(\bm{U}(\bm{\mu}^*)\bm{a}(\bm{\mu^*});\bm{\mu^*})\rVert _p
  \label{MOR10}
\end{equation}
where $\lVert \cdot\rVert _p$ is the standard $\ell_p$-norm. For $p=2$, this residual minimization~\eqref{MOR10} is equivalent to a least-squares Petrov-Galerkin projection of the residual equation in the continuous framework given by~\eqref{MOR8} with test basis $\bm{v}_n=\bm{J}\bm{u}_n$ where $\bm{J}=\partial \bm{R}/\partial \bm{a}$ denotes the Jacobian of the residual.

Similarly, the least-squares minimization problem~\eqref{MOR9} in discrete form can be expressed as:
\begin{equation}
  \underset{\bm{c}^{s}}{\text{min}}
   \sum_{n=1}^{N_v} \left\| \bm{w}'(\bm{\mu}_s,\bm{\mu}_{v_n})-\bm{w}(\bm{\mu}_{v_n})\right\|^2 _2
  \label{MOR11}
\end{equation}

\emph{Remark 1}: Evaluation of the transported snapshot in the discrete form faces a drawback of the possibility of negative volume elements. Since it is usually not possible to evaluate
the solution residual and perform
interpolation on a computational grid with negative volume elements, the training error minimization problem~\eqref{MOR9} can be augmented with inequality constraints for the volume elements:
\begin{equation}
\begin{split}
&   \underset{\bm{c}^{s}}{\text{min}}
   \sum_{n=1}^{N_v} \left\| \bm{w}'(\bm{\mu}_s,\bm{\mu}_{v_n})-\bm{w}(\bm{\mu}_{v_n})\right\|^2 _2 \\ 
& \text{subject to} \quad V_j>\delta_v, \text{ for } j=1,\ldots,N_V
\end{split}
\label{MOR12}
\end{equation}
where $V_j$ are the $N_V$ element volumes and $\delta_v>0$ is the minimal positive
volume. 
For a structured Cartesian grid, this constraint simplifies to
 $(x'_i)_{j+1} -(x'_i)_{j}>\delta_v$ for $i=1,2,3$ and $j =
 1,\ldots,N_V$, where $\bm{x}'_i=\bm{x}_i+\bm{c}_s(\bm{x}_i;\bm{\Delta \mu})$.

\emph{Remark 2}: Since the transported snapshot $\bm{w}'(\bm{\mu}_s,\bm{\mu}_{v_n})$ is an approximation of the snapshot $\bm{w}(\bm{\mu}_{v_n})$, $\bm{w}'(\bm{\mu}_s,\bm{\mu}_{v_n})$ should  satisfy the boundary conditions corresponding to the parameter $\bm{\mu}_{v_n}$. Generally, for a large class of problems, linear extrapolation at the boundaries of the computational domain (during the interpolation of $\bm{w}(\bm{\mu}_{s})$ from $\bm{x}_i+\bm{c}_s(\bm{x}_i;\bm{\Delta \mu})$ to $\bm{x}_i$) are found to be a reasonable approximation of the boundary conditions for low to moderate parameter variations. However, approximation of boundary conditions by extrapolation would be a poor choice if the desired boundary conditions have sharp gradients or discontinuities. To tackle this issue, the transport of the boundary nodes can be restricted such that the need for extrapolation is avoided. The transport of the boundary nodes can be restricted by enforcing appropriate conditions on the coefficients $\bm{c}^{s}$ of the transport $\bm{c}_s(\bm{x}_i;\bm{\Delta \mu})$. Since this issue is problem dependent, it is discussed in detail in \S~\ref{Numericalexperiments} with the help of test problems.

\emph{Remark 3: Choice of norms, $\ell_p$}:
In traditional projection-based MOR, the generalized coordinates are usually
selected to minimize the $\ell_2$-norm of the residual. Although this approach
has been demonstrated to work adequately for many applications, in the case when
the FOM is comprised of a system of hyperbolic conservation laws, minimizing the
$\ell_1$-norm has been shown to be preferable \citep{abgrall2016robust}. For our proposed approach, $\ell_1$ norm for the residual minimization~\eqref{MOR10} is found to perform better than other choices of norms.
Unfortunately, the optimal choice of norm remains an open problem.

The classical approach for solving $\ell_1$-norm minimization problems involves
recasting the problem as a linear program or, alternatively, solving iteratively
using, for example, Iteratively Reweighted Least Squares
(IRLS)~\cite{holland1977robust,daubechies2010iteratively}. Methodology for
minimizing the $\ell_1$-norm is also explained in~\cite{abgrall2016robust}. 


\subsection{Summary of TSMOR}

The offline and online stages of the proposed TSMOR approach in the discrete framework are summarized in Algorithms~\ref{algo1} and~\ref{algo2}, respectively.
\begin{algorithm}
\caption{TSMOR-offline stage}\label{algo1}
\begin{algorithmic}[1]
	\Require Steady state snapshots, $\bm{w}(\bm{\mu_s})$ for $s=1,\dotsc,N_s$
	\Ensure Coefficients, $\bm{c}^{s}$ for $s=1,\dotsc,N_s$
	\For{$s\leftarrow 1$ to $N_s$} 
	\State Determine $N_v$ training snapshots $\bm{w}(\bm{\mu}_{v_n})$ for $n = 1,\ldots,N_v$ as mentioned in \S~\ref{Training_snapshots2}
	\State Determine $\bm{\Delta \mu}= \bm{\mu}_{v_n}-\bm{\mu}_{s}$ for $n = 1,\ldots,N_v$
	\State Select basis functions, $\bm{f}_p(\bm{x})$ and determine $g_q(\bm{\Delta \mu})$ from Eq.~\eqref{distance2}
	\State Define transport fields $\bm{c}_{s}(\bm{x};\bm{\Delta \mu})$ using Eq.~\eqref{transport_polynomial5}
	\State Compute transported snapshots $\bm{w}'(\bm{\mu}_{s},\bm{\mu}_{v_n})$ by interpolating $\bm{w}(\bm{\mu}_{s})$ from $\bm{x}_i+\bm{c}_{s}(\bm{x}_i;\bm{\Delta \mu})$ to the original grid $\bm{x}_i$
	\State Solve training error minimization problem~\eqref{MOR11} or~\eqref{MOR12} for the coefficients $\bm{c}^{s}$
	\EndFor
\end{algorithmic}
\end{algorithm}

\begin{algorithm}
\caption{TSMOR-online stage}\label{algo2}
\begin{algorithmic}[1]
	\Require $\bm{w}(\bm{\mu_s})$ and $\bm{c}^{s}$ for $s=1,\dotsc,N_s$; Basis functions, $\bm{f}_p(\bm{x})$; Unsampled parameter, $\bm{\mu}^*$
	\Ensure Solution prediction, $\bm{w}_r(\bm{\mu}^*)$
	\State Determine $N_k$ local snapshots $\bm{w}(\bm{\mu}_{k_n})$ for $n=1,\dotsc,N_k$ as mentioned in \S~\ref{Basis2}
	\State Determine $\bm{\Delta \mu}= \bm{\mu}^*-\bm{\mu}_{k_n} $ for $n=1,\dotsc,N_k$
	\State Determine $g_q(\bm{\Delta \mu})$ from Eq.~\eqref{distance2}
	\State Compute transports $\bm{c}_{k_n}(\bm{x}_i; \bm{\Delta \mu})$ using Eq.~\eqref{transport_polynomial5}
	\State Compute transported snapshots $\bm{w}'(\bm{\mu}_{k_n},\bm{\mu}^*)$ by interpolating $\bm{w}(\bm{\mu}_{k_n})$ from $\bm{x}_i+\bm{c}_{k_n}(\bm{x}_i;\bm{\Delta \mu})$ to the original grid $\bm{x}_i$
	\State Construct local basis $\bm{U}(\bm{\mu}^*)$ as a collection of transported snapshots, i.e. $\bm{U}(\bm{\mu}^*)_{:,n} := \bm{w}'(\bm{\mu}_{k_n},\bm{\mu}^*)$
	\State Solve the residual minimization problem~\eqref{MOR10} for $\bm{a}(\bm{\mu}^*)$
	\State Compute $\bm{w}_r(\bm{\mu}^*)$ from Eq.~\eqref{MOR44}
\end{algorithmic}
\end{algorithm}


\section{Hyper-reduction}
\label{HR}

In the online stage of TSMOR, evaluation of the residual
$\bm{R}(\bm{U}(\bm{\mu}^*)\bm{a}(\bm{\mu^*}))$ in Eq.~\eqref{MOR10} scales with
the size of the FOM, $N_w$. Furthermore, for every new prediction at $\bm{\mu}^*$, a new parameter dependent basis $\bm{U}(\bm{\mu}^*)$ has to be constructed, the computation of which also scales with $N_w$. Hyper-reduction can significantly reduce these
computational complexities. A review of the state of art hyper-reduction
techniques, such as DEIM~\cite{chaturantabut2010nonlinear},
GNAT~\cite{carlberg2011gnat} and ECSW~\cite{farhat2014dimensional} is provided
in this section. Furthermore, the methodology to equip the proposed TSMOR
approach with hyper-reduction strategies is outlined. 

\subsection{Review of hyper-reduction techniques}

In hyper-reduction for traditional projection-based MOR described in \S~\ref{Traditional}, the residual $\bm{R}(\bm{Ua}(\bm{\mu}^*))$ is evaluated only at a small subset of $n_w$
interpolation entries $\varepsilon \subset \{1,\cdots,N_w\}$. The interpolation
matrix $\bm{Z} \in \mathbb{R}^{N_w \times n_w}$ is thus defined as
\begin{equation}
\bm{Z}=[\bm{e}_{\varepsilon_1},\bm{e}_{\varepsilon_2},\cdots,\bm{e}_{\varepsilon_{n_w}}]
\end{equation}
where $\bm{e}_{\varepsilon} \in \mathbb{R}^{N_w}$ is the $\varepsilon$th canonical unit vector.
$\bm{R}(\bm{Ua}(\bm{\mu}^*))$ is then approximated in a low-dimensional subspace
$\bm{U}_R \in \mathbb{R}^{N_w \times N_R}$
\begin{equation}
\bm{R}(\bm{Ua}(\bm{\mu}^*)) \approx \bm{U}_R (\bm{Z}^T \bm{U}_R)^+ \bm{Z}^T \bm{R}(\bm{Ua}(\bm{\mu}^*))
\label{HR_approx}
\end{equation}
Finally, equation~\eqref{HR_approx} is projected in a low-dimensional
subspace using state basis $\bm{U}$. 
Since the computation of $\bm{Z}^T \bm{R}(\bm{Ua}(\bm{\mu}^*))$ involves the evaluation of the residual at only $n_w$ grid locations, the resulting minimization problem is independent of the size of FOM, $N_w$.

Another class of hyper-reduction techniques, for instance ECSW, involves
minimization of the weighted residuals computed only at $\varepsilon$ points.
Thus, all the computations are performed on these \emph{collocation points} and
interpolation of the residuals using basis functions is avoided.

\subsection{Hyper-reduction applied to TSMOR}
\label{HR2}

In this work, we adopt a collocation-based  hyper-reduction approach similar to ECSW. 
More specifically, the residual
minimization problem~\eqref{MOR10} in this hyper-reduction framework is written
as
\begin{equation}
  \underset{\bm{a}}{\text{min}}
    \lVert \bm{Z}^T \bm{R}(\bm{U}(\bm{\mu}^*)\bm{a}(\bm{\mu^*});\bm{\mu^*})\rVert _p
  \label{MOR13}
\end{equation}

Generally, in most CFD problems, the Jacobian matrix is sparse. Hence the
computation of the residuals is dependent only on a few subset of $\hat{n}_w$
entries $\hat{\varepsilon} \subset \{1,\cdots,N_w\}$. The corresponding
interpolation matrix is denoted as $\bm{P} \in \mathbb{R}^{N_w \times \hat{n}_w}$. Thus $\bm{U}(\bm{\mu^*}) \bm{a}(\bm{\mu^*})$ can be
replaced by $\bm{P} \bm{P}^T \bm{U}(\bm{\mu^*}) \bm{a}(\bm{\mu^*})$ resulting in the cheap computation of the collocated reduced-basis, $\bm{P}^T \bm{U}(\bm{\mu^*})$, whose columns contain the collocated transported snapshots, i.e., $(\bm{P}^T \bm{U}(\bm{\mu^*}))_{:,n}=\bm{P}^T \bm{w}'(\bm{\mu}_{k_n},\bm{\mu}^*)$. 

The collocated transported snapshot at the collocation points $\hat{\varepsilon}$ is
computed by interpolating the collocated snapshot, $\bm{P}^T \bm{w}(\bm{\mu}_{k_n})$, from $\bm{\hat{x}}_i+\bm{\hat{c}}_{k_n}(\bm{\hat{x}}_i;\bm{\Delta \mu})$ to the original grid $\bm{\hat{x}}_i$. 
Here $\bm{\hat{c}}_{k_n}(\bm{\hat{x}}_i;\bm{\Delta \mu})=\bm{P}^T \bm{{c}}_{k_n}(\bm{x}_i;\bm{\Delta \mu}) \in \mathbb{R}^{\hat{n}_w}$ are the collocated transports and $\bm{\hat{x}}_i= \bm{P}^T \bm{x}_i \in \mathbb{R}^{\hat{n}_w}$ are the Cartesian coordinates of the original grid at the collocation points.  

To summarize, equation~\eqref{MOR13} involves the computation of $\bm{Z}^T \bm{R}(\bm{P P}^T \bm{U}(\bm{\mu}^*)\bm{a}(\bm{\mu^*}))$ which necessitates the computation of $\bm{R}(\bm{U}(\bm{\mu}^*)\bm{a}(\bm{\mu^*}))$ and $\bm{U}(\bm{\mu}^*)$ only at $\varepsilon$ and $\hat{\varepsilon}$ indices respectively, resulting in a reduction of the computational complexity from $N_w$ to $\hat{n}_w$.

\subsubsection{Identification of collocation points}

Various hyper-reduction techniques in literature employ different strategies for
identifying the collocation points or interpolation entries $\varepsilon$.
Generally, these approaches are specific to their respective hyper-reduction
procedure. For instance, these algorithms are based on minimization of the error
in the interpolated snapshots~\cite{chaturantabut2010nonlinear}, greedy approach
to minimize error associated with gappy-POD projection of
residual~\cite{carlberg2013gnat} and solving a sparse non-negative least-squares
(NNLS) problem~\cite{farhat2014dimensional}. 
In this work, we employ the standard DEIM Algorithm~\ref{algo3} to identify
the collocation points.

The DEIM algorithm uses basis functions $\bm{U}_R$ of the
nonlinear residual to identify the interpolation entries $\varepsilon$. In our
implementation of the algorithm, $\bm{U}_R$ are the POD basis
of the snapshots of residuals. The snapshots of residuals are collected at each iteration while solving the FOM Eq.~\eqref{steady1} during the offline stage.
For boundary value problems, in addition to
DEIM indices, it is important to include inlet/outlet grid points into
$\varepsilon$ since these boundary conditions contain vital information about
the dynamics of the problem. Details about inclusion of these boundary points are explained in \S~\ref{Numericalexperiments} where this topic is covered for each flow problem. Finally, corresponding interpolation entries
$\hat{\varepsilon}$ for computing the residuals can be related to $\varepsilon$
depending on the type of finite-difference/volume/element scheme.


\begin{algorithm}
  \caption{DEIM}\label{algo3}
  \begin{algorithmic}[1]
    \Require POD basis of snapshots of residuals, $\bm{U}_R$
    \Ensure Interpolation entries $\bm{\varepsilon}=[\varepsilon_1,\cdots,\varepsilon_{n_w}]$ 
    \State $[~,\varepsilon_1]=\verb=max=\{|\bm{U}_{R_{(:,l)}}|\}$
    \State $\bm{V}=[\bm{U}_{R_{(:,l)}}]$, $\bm{Z}=[\bm{e}_{\varepsilon_1}]$, $\bm{\varepsilon}=[\varepsilon_1]$
    \For{$l\leftarrow 2$ to $n_w$}
    \State Solve $(\bm{Z}^T\bm{V})c=\bm{P}^T \bm{U}_{R_{(:,l)}}$
    \State $\bm{r}=\bm{U}_{R_{(:,l)}}-\bm{Vc}$
    \State $[~, \varepsilon_l]=\verb=max=\{|\bm{r}|\}$
    \State $\bm{V}\leftarrow[\bm{V},\bm{U}_{R_{(:,l)}}]$, $\bm{Z}\leftarrow[\bm{Z},\bm{e}_{\varepsilon_l}]$, $\bm{\varepsilon}\leftarrow[\bm{\varepsilon},\varepsilon_l]$
    \EndFor
  \end{algorithmic}
\end{algorithm}

\section{Comparison with previous MOR approaches} 
\label{Comparison}

Our proposed approach shares similarities with previous parametric MOR
techniques in the literature. Similar to Reiss et al.~\cite{reiss2018shifted}, Iollo and Lombardi~\cite{iollo2014advection} and Welper~\cite{welper2017interpolation}, we approximate the sought after solutions
as a superposition of transported snapshots or their basis. However, the methodology to identify the transport fields and corresponding transported snapshots is different where our approach provides the following advantages. 

Shifted-POD method~\cite{reiss2018shifted}, developed for unsteady flow problems, involves the identification of shift velocities based on known or data-driven unsteady transport phenomena. However, this approach has not yet been realized and demonstrated for the reduction of steady flow problems. 

The transport obtained by solving an optimal mass transfer problem~\cite{iollo2014advection} is optimal in the sense that the displacement of the computational domain to transform one snapshot to the other is minimal. The formulation of the optimal mass transfer problem, however, has the following issues. Firstly, the optimal transports obtained between various pairs of snapshots may not provide an intermediate transported snapshot in the predictive regime that approximately satisfies the physics of the underlying problem. In other words, we seek a physics-based transport field instead of an optimal transport (though an optimal transport may show similarities with physics based transport for a subset of flow problems). In our proposed approach, the transport field is evaluated by a least-squares fitting procedure which minimizes the approximation error of multiple local transported snapshots. This ensures that the transport field captures the physics-based dynamics of snapshot transformation from all the neighboring snapshots. Secondly, it has been demonstrated for image processing applications that for images with sharp features, optimal transport can lead to numerical artifacts at these sharp edges~\cite{papadakis2015optimal}, while a Lagrangian approach for solving the optimal mass transfer problem can lead to diffusion of sharp boundaries~\cite{iollo2011lagrangian}. In the fluid mechanics community, these sharp features can be interpreted as shocks, implying that optimal transport may not be easily extendable for flow problems containing shocks.

In transformed snapshot interpolation (TSI)~\cite{welper2017interpolation}, the transports for approximating the solution in the predictive regime are obtained by interpolating the transports evaluated between various pairs of known snapshots. Our proposed method for identifying the transport fields can be considered as a generalization of TSI, with the difference being that our method is essentially a least-squares fit in contrast to interpolation in TSI. The advantage of fitting as compared to interpolation is that the fitting polynomial can be chosen to contain fewer terms than a corresponding interpolation polynomial. This property allows for easy extension of our method to multi-dimensional parameter variations by mitigating the curse of dimensionality faced by TSI. Furthermore, our residual minimization approach is projection-based in contrast to the interpolation-based approach in TSI. One of the potential advantages of projection-based methods is that they retain the
underlying structure of the dynamical system and thus provide, in principle,
more robust predictive capabilities.    

Finally, similar
to Abgrall et al.~\cite{abgrall2016robust}, we minimize the system residual in the
$\ell_1$ norm which has been shown to be beneficial for problems containing
shocks and discontinuities.


\section{Numerical experiments}
\label{Numericalexperiments}

In this section, TSMOR is applied to the steady Euler equations modeling
supersonic flow inside a quasi 1-D nozzle and 2-D flow over a forward facing
step and a nonlinear advection-diffusion equation modeling a jet diffusion flame in a
combustor. These problems are chosen because the steady flow solutions contain
shocks or flame fronts whose spatial locations and orientations are parameter
dependent.  Similar numerical experiments have been studied
by Lucia et al.~\cite{lucia2001reduced}, Mojgani and Balajewicz~\cite{mojgani2017lagrangian}, Zahr et al.~\cite{zahr2013construction}; Welper~\cite{welper2017h}; and Galbally et al.~\cite{galbally2010non} respectively. 

In this
section, the results generated by the proposed TSMOR approach are compared
with traditional projection-based MOR techniques such as LSPG with test subspace $\bm{v}_n =
\bm{Ju}_n$ as described in \S~\ref{Traditional} where $\bm{u}_n$ is the POD basis of the snapshot matrix and $\bm{J}$ is the Jacobian of the residual function. Furthermore, comparison is also made with recent parametric MOR techniques such as $L_1$-dictionary approach~\cite{abgrall2016robust} where the solution is given by a linear combination of local snapshots or \emph{dictionary} elements and $L_1$-norm of the residual is minimized. All results considered in this section are predictive, that is, the predicted
solutions all lie in parameter regions not sampled during the offline training
phase. The performance of these MOR techniques are analyzed by computing the relative error between the predicted and FOM solutions where the error is defined as:
\begin{equation}
Error(\%)=\frac{\lvert| \bm{w}(\bm{\mu}^*)-\bm{w}_r(\bm{\mu}^*)  \rvert|_2}{\lvert|\bm{w}(\bm{\mu}^*)\rvert|_2} \times 100
\end{equation}
where $\bm{w}(\bm{\mu}^*)$ and $\bm{w}_r(\bm{\mu}^*)$ are the FOM and predicted solutions using the above-mentioned MOR methods, respectively. All the computations were done in Matlab.

\subsection{Quasi 1-D flow in a converging-diverging nozzle}
\subsubsection{Problem description}
\label{problemdescription1}
The 1-D Euler equations in a quasi 1-D converging-diverging nozzle
are considered:
\begin{equation}
  \frac{1}{A}\frac{\partial A\bm{F}}{\partial x}=\bm{Q} \hspace{1cm} x\in[0,L]
\label{governing1}
\end{equation}
where $A=A(x)$ is the area profile and 
 $$ \bm{w}(\mu)=\begin{bmatrix}
  \rho\\ 
  \rho u\\ 
  \rho E
  \end{bmatrix}, \quad
  \bm{F}=\begin{bmatrix}
  \rho u\\ 
  \rho u^2+p\\ 
  (\rho E+p)u
  \end{bmatrix}, \quad
  \bm{Q}=\begin{bmatrix}
  0\\ 
  \frac{p}{A}\frac{\partial A}{\partial x}\\ 
  0
  \end{bmatrix}$$
with homogeneous Dirichlet boundary conditions $\rho(0;\mu)=1, \ p(0;\mu)=1$ and
$p(L;\mu)=0.7$.

The boundary conditions are chosen such that a shock is formed in the diverging
section of the nozzle. Length of the nozzle $L$ is 10 units. Area profile of the
converging-diverging nozzle is parabolic with equal inlet and outlet area, $A(0)=A(L)=3$, and
the throat is located at $L/2$. For this problem, the throat area $\mu=A(L/2)$
is the parameter of interest. Steady state solutions are obtained by
discretizing the corresponding governing equations in space using a central
second-order finite difference scheme on a uniform grid which is divided into 1000 grid points with grid spacing
$\Delta x=0.01$. A first-order accurate artificial viscosity scheme using
$\nu=\Delta x / 2$ is used to stabilize the solution. The resulting nonlinear
system of algebraic equations is solved in Matlab using the built-in \verb=fsolve=
algorithm. Fig.~\ref{4sol1} shows the steady density solutions for different
values of throat area $\mu$.

\subsubsection{Implementation of TSMOR}
\label{implementation1}
A snapshot matrix $\bm{M}$ containing 4 snapshots at parameters $\mu_s=[0.5, 0.875, 1.25, 1.625]$ is generated. The coefficients of the polynomial expansion~\eqref{transport_polynomial3} $\bm{c}^{s}$ for each snapshot are computed offline by solving the training error minimization~\eqref{MOR12}. The basis $f_p(x)$ are chosen to be Fourier sine series with $m$ modes:
\begin{equation}
\bm{f}(x)=\left[1, \sin\left(\frac{\pi {x}}{L}\right), \ldots, \sin\left(\frac{(m-1)\pi {x}}{L}\right) \right]
\label{cx}
\end{equation}
For this simple problem, $g_q(\Delta \mu)$  is given by: 
\begin{equation}
g(\Delta \mu)=\Delta \mu
\end{equation}
The interpolation from the
transported grid to the original grid for calculating the transported snapshots
was performed using \verb=interp1= algorithm. The training error minimization~\eqref{MOR12} is solved using \verb=fmincon= algorithm.

\begin{figure}[h]
\centering
\begin{subfigure}[t]{0.4\textwidth}
\centering
\includegraphics[scale=1]{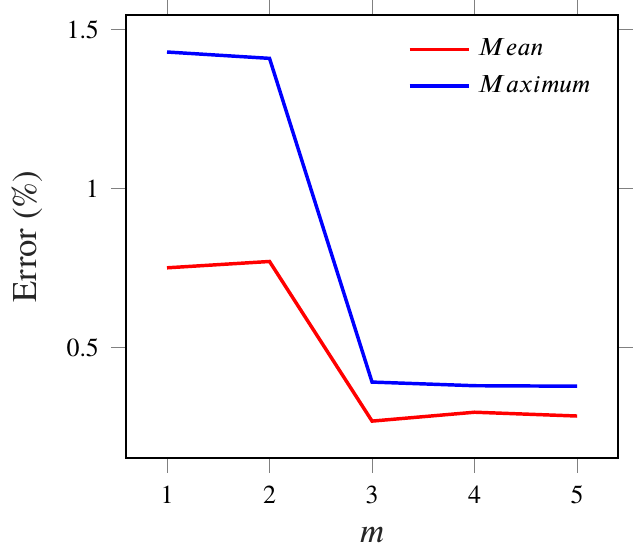}
\caption{Plot of relative error v/s number of Fourier modes $m$ for TSMOR predicted solutions}
\label{nozzle_Fourier_conv1}
\end{subfigure}\hspace{1cm}
\begin{subfigure}[t]{0.4\textwidth}
\centering
\includegraphics[scale=1]{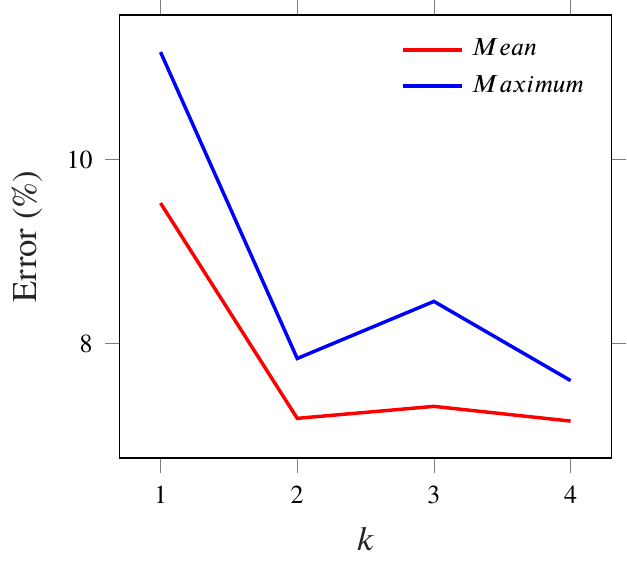}
\caption{Plot of relative error v/s number of POD basis $k$ for LSPG predicted solutions}
\label{nozzle_Fourier_conv2}
\end{subfigure}
\caption{Convergence plot of relative error for TSMOR and LSPG predicted solutions}
\label{nozzle_Fourier_conv}
\end{figure}

First, convergence of the proposed TSMOR approach with respect to the number of Fourier modes $m$ is studied by predicting new solutions in the predictive regime $\mu^*$. Fig.~\ref{nozzle_Fourier_conv1} shows the mean and maximum relative error in the TSMOR solutions predicted at two uniformly distributed parameters in every interval of $\mu_s$ for different number of Fourier modes $m$. It can be seen that as the number of Fourier modes increases, the error converges to a low value of 0.27\%. The TSMOR convergence plot is compared to the convergence of LSPG solutions with respect to the number of POD basis $k$ of the snapshot matrix $\bm{M}$. Similar to Fig.~\ref{nozzle_Fourier_conv1}, Fig.~\ref{nozzle_Fourier_conv2} displays the relative errors in the LSPG solutions predicted at the same set of parameters for different number of bases. It can be observed that, the error converges only to 7.16\% even though all 4 POD bases were used for prediction. Hereafter, all the TSMOR predicted results presented for this problem are produced with 3 Fourier modes.
\begin{figure}[h]
\centering
\begin{subfigure}[t]{0.4\textwidth}
\centering
\includegraphics[scale=1]{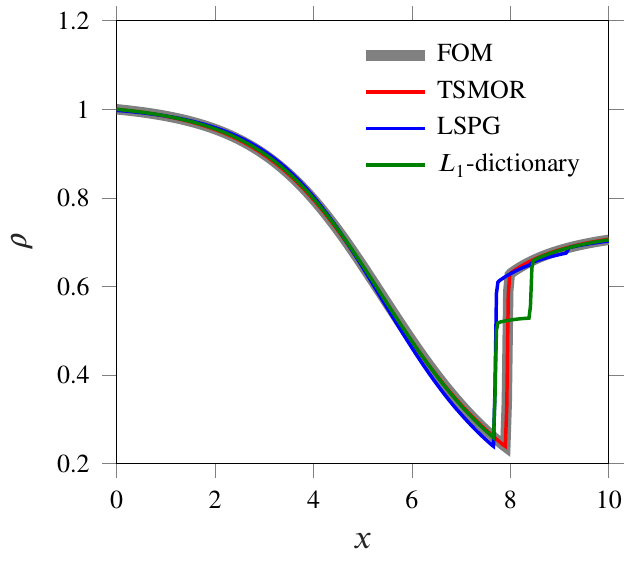}
\caption{Predicted solutions at $\mu^*=1.00$}
\label{nozzle_predict1}
\end{subfigure}\hspace{1cm}
\begin{subfigure}[t]{0.4\textwidth}
\centering
\includegraphics[scale=1]{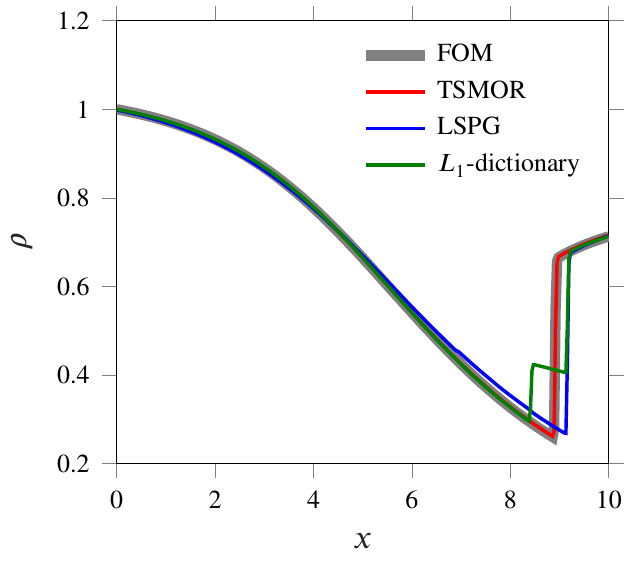}
\caption{Predicted solutions at $\mu^*=1.50$}
\label{nozzle_predict2}
\end{subfigure}
\caption{Comparison of predicted solutions using TSMOR, LSPG and $L_1$-dictionary approach with FOM}
\label{nozzle_predict}
\end{figure}

Next, the performance of the proposed TSMOR approach is compared to several existing MOR techniques. Fig.~\ref{nozzle_predict} illustrates the predictive capabilities of several MOR
approaches for the parameters $\mu^*=1.0$ and $\mu^*=1.5$. The FOM density solution is given by the gray
lines while the new proposed TSMOR approach using 2 local bases corresponding to two nearby snapshots is given by the red lines. The blue lines correspond to the solution obtained by LSPG method using 4 POD modes of the snapshot matrix $\bm{M}$. The green lines correspond to $L_1$-dictionary approach using 2 local snapshots or dictionary elements. The proposed TSMOR
approach reproduces the solution remarkably well. In contrast, LSPG does not predict the correct shock
location while $L_1$-dictionary solutions are dominated by staircase shock type errors.

\begin{figure}[h]
\centering
\begin{subfigure}[t]{0.4\textwidth}
\centering
\includegraphics[scale=1]{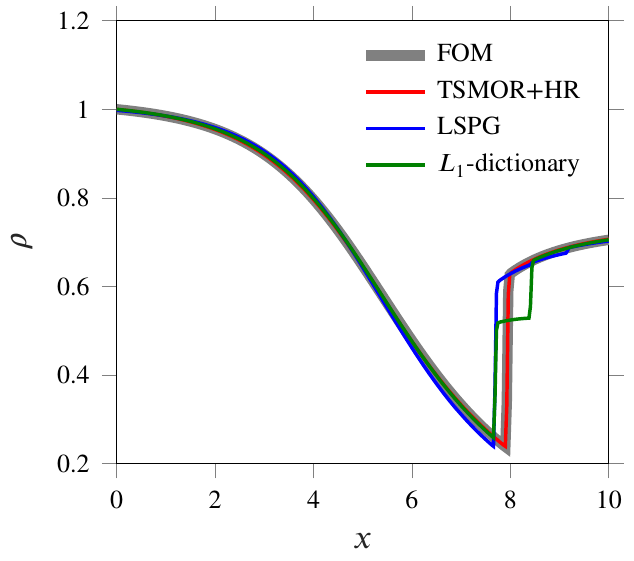}
\caption{Predicted solutions at $\mu^*=1.00$}
\label{nozzle_predictHR1}
\end{subfigure}\hspace{1cm}
\begin{subfigure}[t]{0.4\textwidth}
\centering
\includegraphics[scale=1]{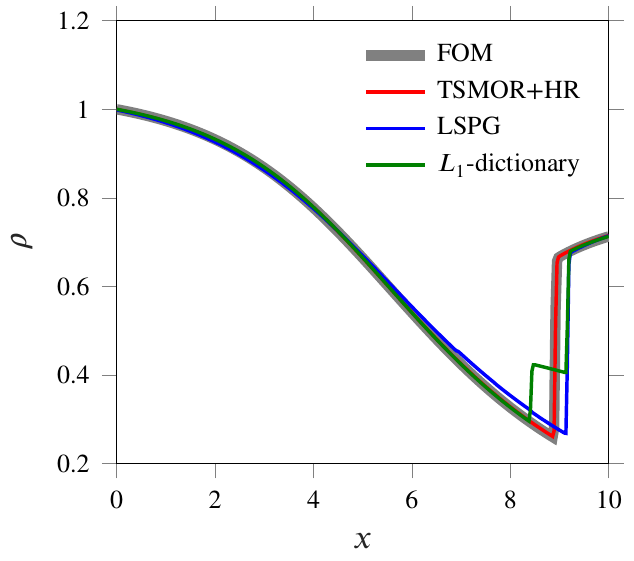}
\caption{Predicted solutions at $\mu^*=1.50$}
\label{nozzle_predictHR2}
\end{subfigure}
\caption{Comparison of predicted solutions using hyper-reduced TSMOR, LSPG and $L_1$-dictionary approach with FOM}
\label{nozzle_predictHR}
\end{figure}

Next, the TSMOR approach is equipped with the hyper-reduction strategy mentioned in \S~\ref{HR2}. 
First, 30 collocation points are
obtained by employing the DEIM algorithm. Second, these points are augmented
with inlet and outlet points, $i=1$, and $i=N_w$, respectively, if not already included in those 30 DEIM collocation points.
Finally, an additional $\hat{n}_w\approx n_w \times 2=64$
points $\hat{\varepsilon}$ are included to enable the evaluation of the residuals via
the central finite difference scheme. 
Fig.~\ref{nozzle_predictHR} illustrates the predictive capabilities of hyper-reduced TSMOR (TSMOR+HR)
for the same parameters $\mu^*=1.0$ and $\mu^*=1.5$. 
The hyper-reduced TSMOR density solutions are compared with non-hyper-reduced LSPG and $L_1$-dictionary approaches. As before, excellent agreement with the FOM solution is demonstrated by hyper-reduced TSMOR.

\begin{figure}
\centering
\includegraphics[scale=1]{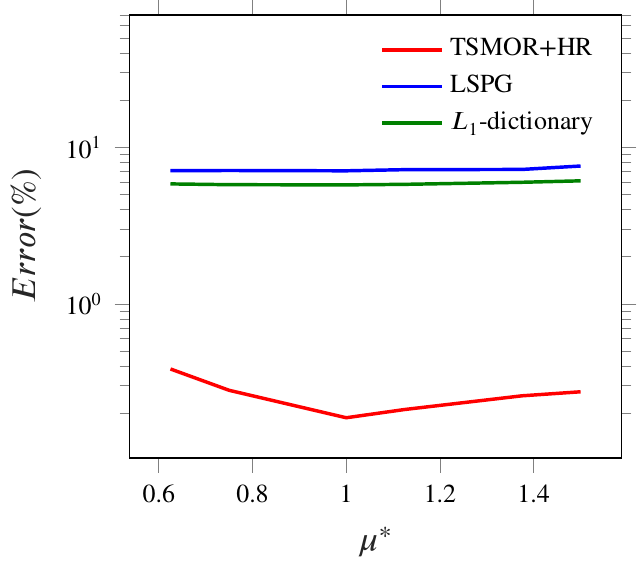}
\caption{Performance comparison between hyper-reduced TSMOR, LSPG and $L_1$-dictionary approach for solution predictions at various parameters}
\label{nozzle_error}
\end{figure}

\begin{figure}
\centering
\begin{subfigure}[t]{0.4\textwidth}
\centering
\includegraphics[scale=1]{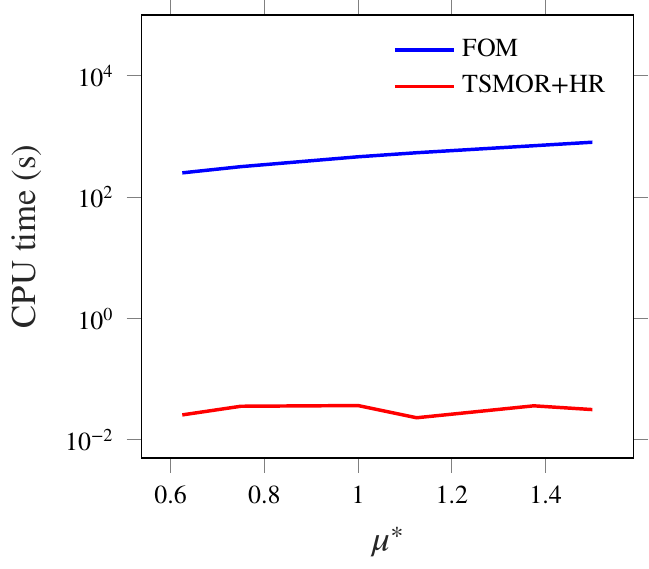}
\caption{CPU wall-time}
\label{nozzle_time1}
\end{subfigure}\hspace{1cm}
\begin{subfigure}[t]{0.4\textwidth}
\centering
\includegraphics[scale=1]{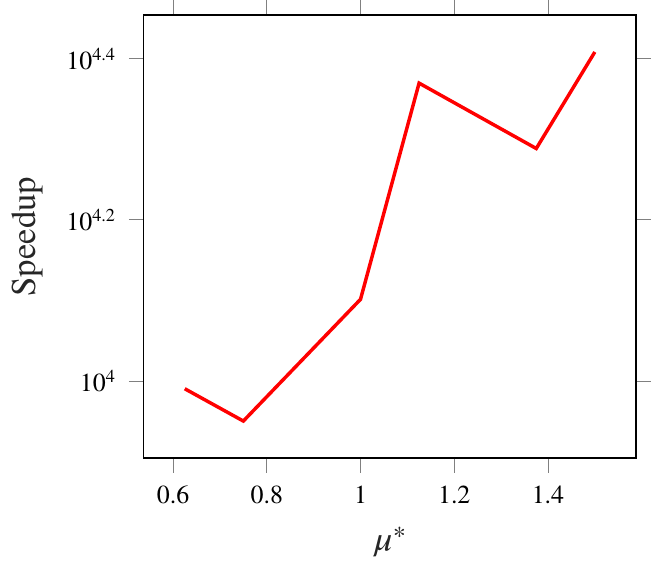}
\caption{Speedup achieved by hyper-reduced TSMOR}
\label{nozzle_time2}
\end{subfigure}
\caption{Comparison of CPU wall-time associated with FOM and online stage of hyper-reduced TSMOR for solution predictions at various parameters}
\label{nozzle_time}
\end{figure}

In Fig.~\ref{nozzle_error}, relative errors between the FOM solutions and
predicted solutions using hyper-reduced TSMOR, non-hyper-reduced LSPG and $L_1$-dictionary approaches are illustrated across the entire parameter range of interest.  For
this case, predictions are made at two uniformly distributed parameters in every interval of $\mu_s$. It can be observed that the solutions predicted using hyper-reduced TSMOR have an average error of only $0.27\%$ as compared to $7.22\%$ in LSPG and $5.88\%$ in $L_1$-dictionary approach. Thus, for all parameters considered, the TSMOR approach significantly
outperforms LSPG and $L_1$-dictionary methods.  
Finally, wall-times and speed-ups for the FOM and the online stage of hyper-reduced TSMOR are illustrated in
Fig.~\ref{nozzle_time1} and Fig.~\ref{nozzle_time2}, respectively. Here, speed-up is defined as the ratio of wall-times of FOM to the online stage of hyper-reduced TSMOR. TSMOR+HR
delivers a speed-up of approximately four orders of magnitude across the entire
parameter range while maintaining a high level of accuracy.

\subsection{Supersonic flow over a forward facing step}
\subsubsection{Problem description}

\begin{figure}[h]
\centering
\begin{subfigure}[b]{1\textwidth}
\centering
\includegraphics[scale=1]{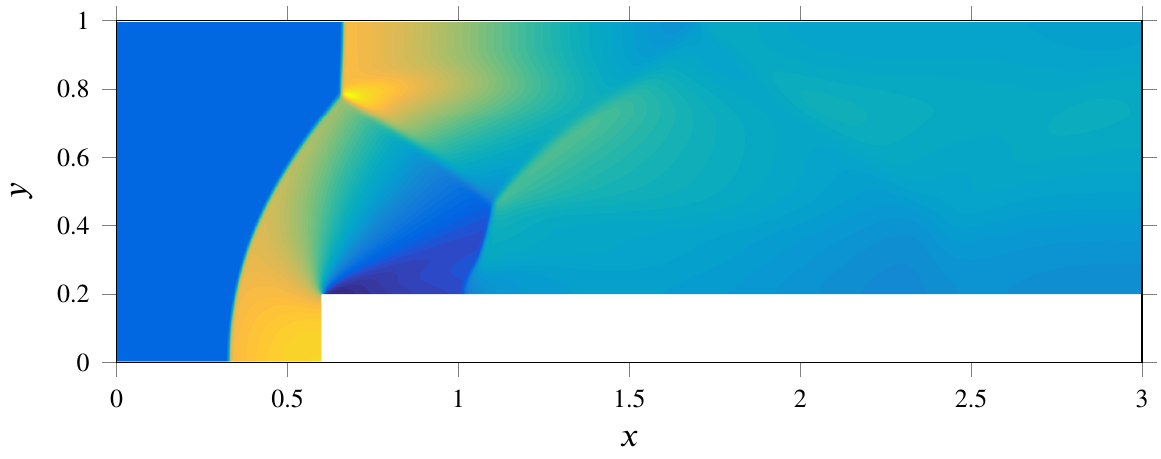}
\caption{Steady density contour for inlet Mach number $\mu=3.3$}
\label{known1}
\end{subfigure}\vspace{0.5cm}
\begin{subfigure}[b]{1\textwidth}
\centering
\includegraphics[scale=1]{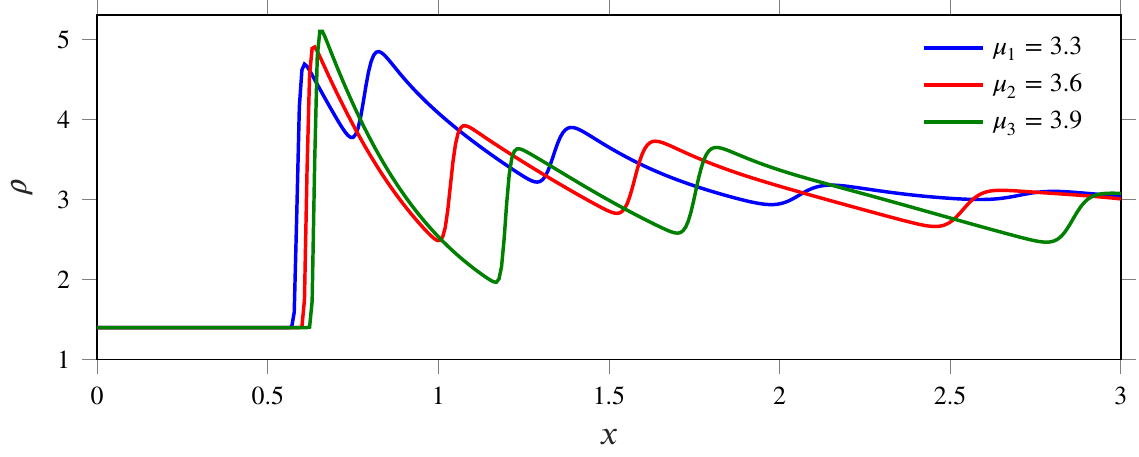}
\caption{1-D steady density plots at $y=0.7$ for inlet Mach number $\mu$}
\label{known2}
\end{subfigure}
\caption{Steady state density plots of a supersonic flow over a forward facing step}
\label{known}
\end{figure}

This problem consists of a supersonic flow over a 2-D forward facing step in a
wind tunnel setup with walls on top and bottom as described
by~\cite{woodward1984numerical} and also shown in Fig.~\ref{known1}. Length (L)
and height (H) of the wind tunnel are 3 units and 1 unit respectively. The step
has a height of 0.2 units and is located at the bottom wall starting at 0.6
units from the left-end of the tunnel. 2-D Euler equations governing the
supersonic flow over a forward facing step are:
\begin{equation}
\frac{\partial \bm{F}}{\partial x}+\frac{\partial \bm{G}}{\partial y}=0, \hspace{1cm} x\in[0,L], \hspace{0.3cm} y\in[0,H]
\label{governing3}
\end{equation}
where
$$\bm{w}(\mu)=\begin{bmatrix}
\rho\\ 
\rho u\\ 
\rho v\\
\rho E
\end{bmatrix};
\bm{F}=\begin{bmatrix}
\rho u\\ 
\rho u^2+p\\ 
\rho uv\\
(\rho E+p)u
\end{bmatrix};
\bm{G}=\begin{bmatrix}
\rho v\\ 
\rho uv\\
\rho v^2+p\\ 
(\rho E+p)v
\end{bmatrix}$$
with homogeneous Dirichlet inlet boundary conditions $\rho(0,y;\mu)=1.4, \
p(0,y;\mu)=1, \ u(0,y;\mu)=\mu$ and $v(0,y;\mu)~=~0$;
where $\mu$ is the inlet supersonic Mach number which is taken to be the varying
parameter of interest. No penetration solid wall boundary conditions are imposed
on the top, bottom and step wall surfaces. The equations are discretized in
space using a second-order, central finite difference scheme on a uniform
Cartesian grid which is divided into 0.48 million grid points with $\Delta x=\Delta
y=0.025$. The solutions are stabilized using a first-order artificial viscosity
scheme where the artificial viscosity is set to be $\nu=\Delta x / 0.8$. The
resulting equations are solved by marching to steady state using a second-order
Strong Stability Preserving (SSP) Runge--Kutta scheme~\cite{gottlieb2005high}.
Fig.~\ref{known1} shows the steady state density contour for inlet Mach number $\mu=3.3$,
while the corresponding 1-D plots at $y=0.7$ for inlet Mach numbers $\mu=3.3$, $3.6$, and $3.9$ are shown in Fig.~\ref{known2}. 

\subsubsection{Implementation of TSMOR}

\begin{figure}[h]
\centering
\begin{subfigure}[t]{0.4\textwidth}
\centering
\includegraphics[scale=1]{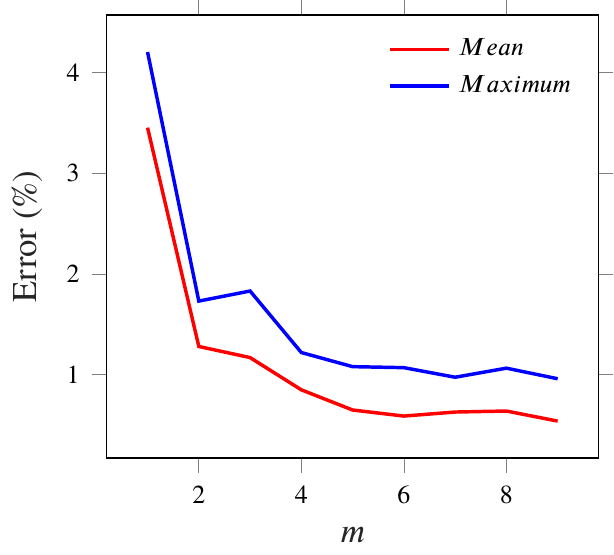}
\caption{Plot of relative error v/s number of Fourier modes $m$ for TSMOR predicted solutions}
\label{step_Fourier_conv1}
\end{subfigure}\hspace{1cm}
\begin{subfigure}[t]{0.4\textwidth}
\centering
\includegraphics[scale=1]{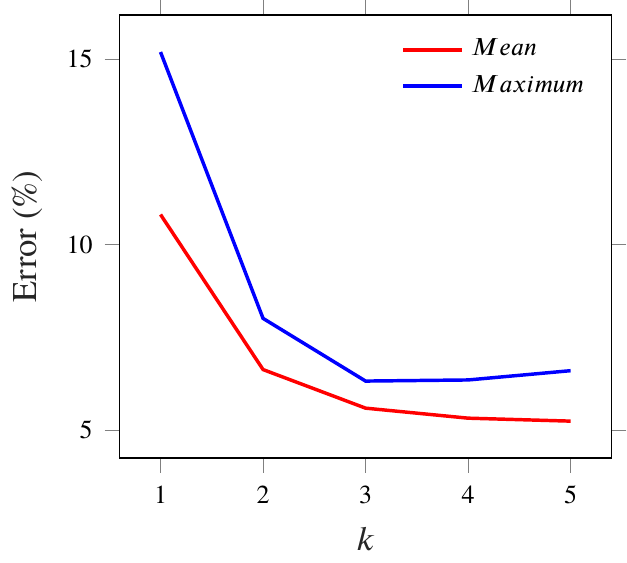}
\caption{Plot of relative error v/s number of POD basis $k$ for LSPG predicted solutions}
\label{step_Fourier_conv2}
\end{subfigure}
\caption{Convergence plot of relative error for TSMOR and LSPG predicted solutions}
\label{step_Fourier_conv}
\end{figure}

A snapshot matrix $\bm{M}$ containing 5 snapshots at parameters $\mu_s=[3.3, 3.45, 3.6, 3.75, 3.9]$ is generated. 
The coefficients of the polynomial expansion~\eqref{transport_polynomial5} $\bm{c}^{s}$ for each snapshot are computed offline by solving the training error minimization~\eqref{MOR12}. The bases of the polynomial expansion, $f_{x_p}(x)$ and $f_{y_p}(y)$, for the $x$ and $y$ transport fields, $c_{s_x}(x,\Delta \mu)$ and $c_{s_y}(y,\Delta \mu)$, respectively, are chosen to be Fourier sine series with $m$ modes each:
\begin{equation}
\begin{gathered}
\bm{f}_x(x)=\left[1, \sin\left(\frac{\pi {x}}{L}\right), \ldots, \sin\left(\frac{(m-1)\pi {x}}{L}\right) \right]\\
\bm{f}_y(y)=\left[1, \sin\left(\frac{\pi {y}}{H}\right), \ldots, \sin\left(\frac{(m-1)\pi {y}}{H}\right) \right]
\end{gathered}
\label{cx2}
\end{equation}
and $g_q(\Delta \mu)$ is given by:
\begin{equation}
\bm{g}(\Delta \mu)=\left[\Delta \mu, \Delta \mu^2 \right]
\end{equation}
The interpolation from the
transported grid ($\bm{x}+\bm{c}_{s_x}(\bm{x};{\Delta \mu}),\bm{y}+\bm{c}_{s_y}(\bm{y};{\Delta \mu})$) to the original grid ($\bm{x},\bm{y}$) for calculating the transported snapshots
was performed using \verb=interp2= algorithm in Matlab. The training error minimization~\eqref{MOR12} is solved using the \verb=fmincon= algorithm.

First, convergence of the proposed TSMOR approach with respect to the number of Fourier modes $m$ is studied by predicting new solutions in the predictive regime $\mu^*$. Fig.~\ref{step_Fourier_conv1} shows the mean and maximum relative error in the TSMOR solutions predicted at two uniformly distributed parameters in every interval of $\mu_s$ for different number of Fourier modes $m$. It can be seen that as the number of Fourier modes increases, the error converges to a low value of 0.54\%. The TSMOR convergence plot is compared to the convergence of LSPG approach with respect to the number of POD basis $k$ of the snapshot matrix $\bm{M}$. Similar to Fig.~\ref{step_Fourier_conv1}, Fig.~\ref{step_Fourier_conv2} displays the relative errors in the LSPG solutions predicted at the same set of parameters for different number of bases. It can be observed that, the error converges only to 5.24\% even though all 5 POD basis were used for prediction. Hereafter, all the TSMOR predicted results presented for this problem are produced with 9 Fourier modes.

\begin{figure}[h!]
\centering
\begin{subfigure}[b]{1\textwidth}
\centering
\includegraphics[scale=1]{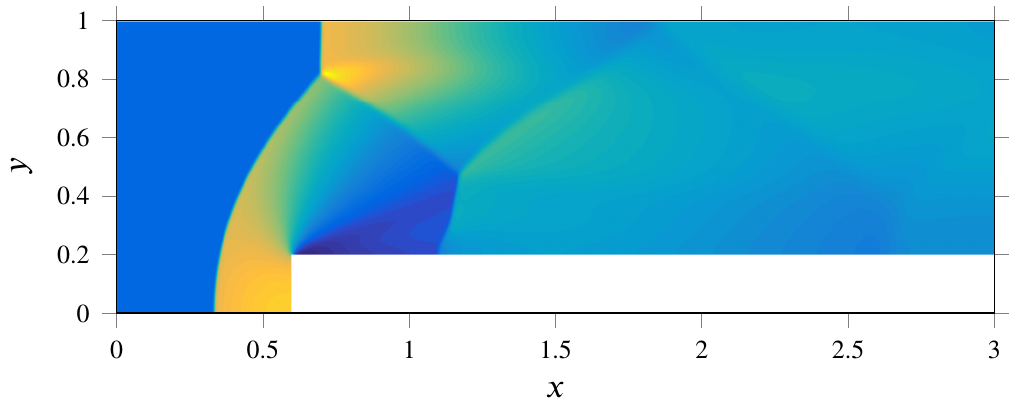}
\caption{FOM solution}
\label{step_FOM}
\end{subfigure}\vspace{0.75cm}
\begin{subfigure}[b]{1\textwidth}
\centering
\includegraphics[scale=1]{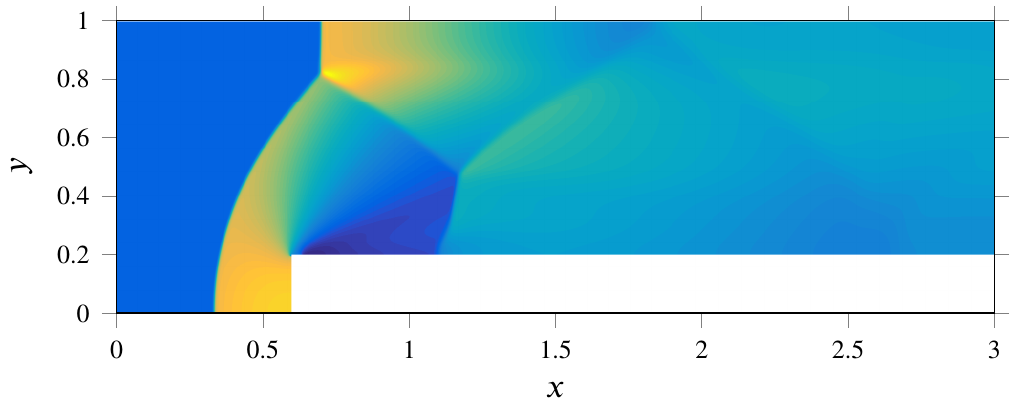}
\caption{TSMOR solution}
\label{step_TSMOR}
\end{subfigure}\vspace{0.75cm}
\begin{subfigure}[b]{1\textwidth}
\centering
\includegraphics[scale=1]{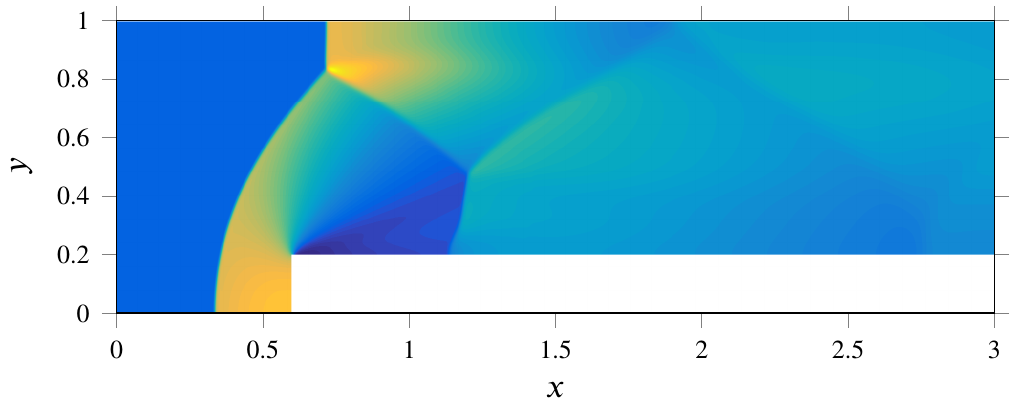}
\caption{LSPG solution}
\label{step_LSPG}
\end{subfigure}\vspace{0.75cm}
\begin{subfigure}[b]{1\textwidth}
\centering
\includegraphics[scale=1]{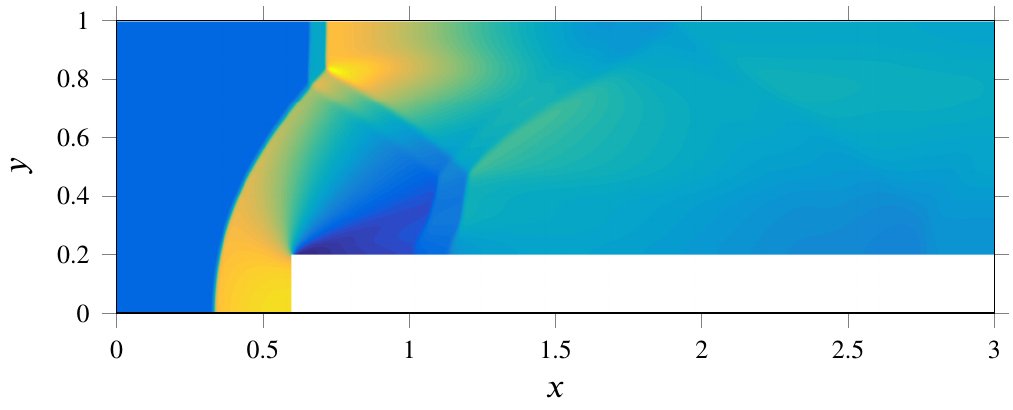}
\caption{$L_1$-dictionary solution}
\label{step_L1}
\end{subfigure}
\caption{Comparison of steady state density solutions at $\mu^*=3.4$ predicted
by TSMOR, LSPG and $L_1$-dictionary approaches with FOM solution}
\label{step_pred}
\end{figure}

\begin{figure}[htbp!]
\centering
\begin{subfigure}{1\textwidth}
\centering
\includegraphics[scale=1]{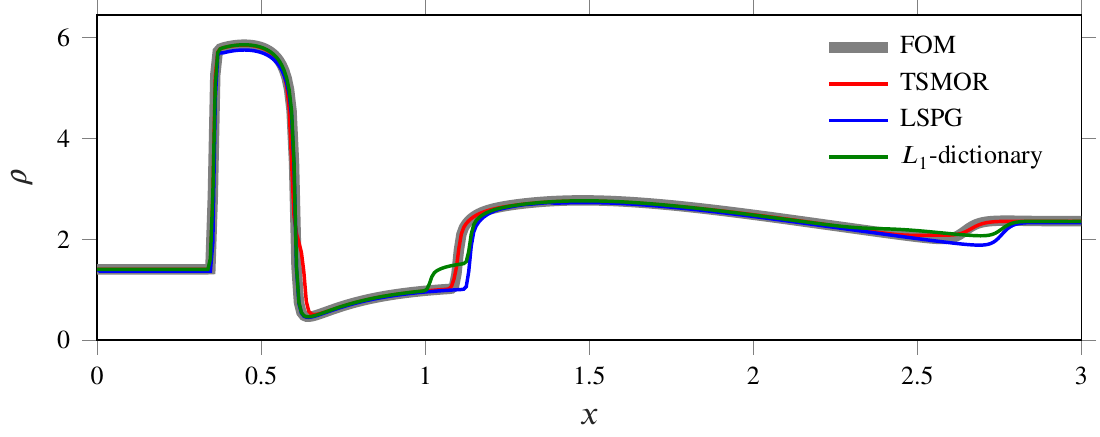}
\caption{$y=0.205$}
\end{subfigure}\vspace{0.75cm}
\begin{subfigure}{1\textwidth}
\centering
\includegraphics[scale=1]{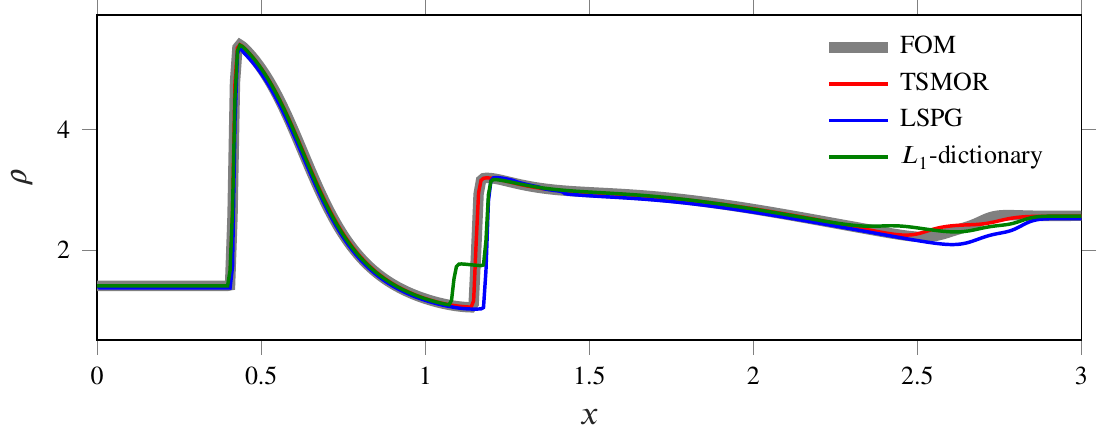}
\caption{$y=0.4$}
\end{subfigure}\vspace{0.75cm}
\begin{subfigure}{1\textwidth}
\centering
\includegraphics[scale=1]{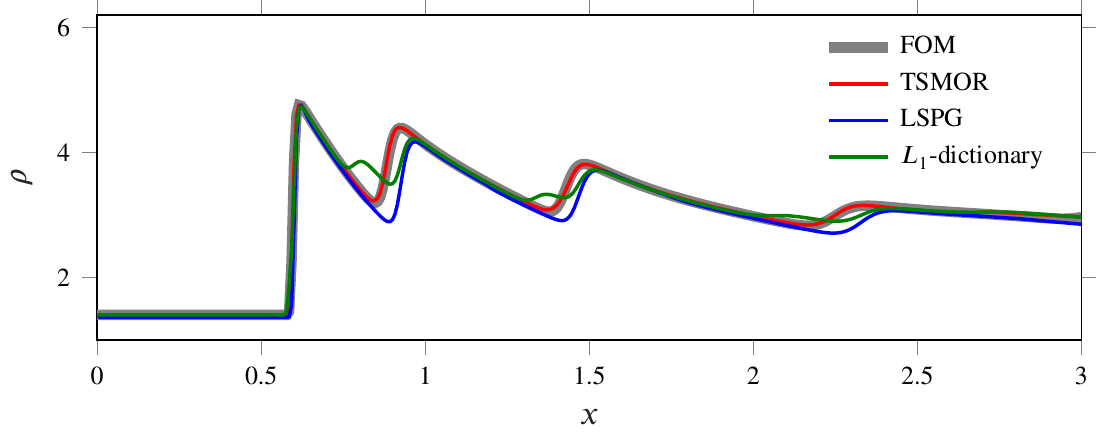}
\caption{$y=0.7$}
\end{subfigure}\vspace{0.75cm}
\begin{subfigure}{1\textwidth}
\centering
\includegraphics[scale=1]{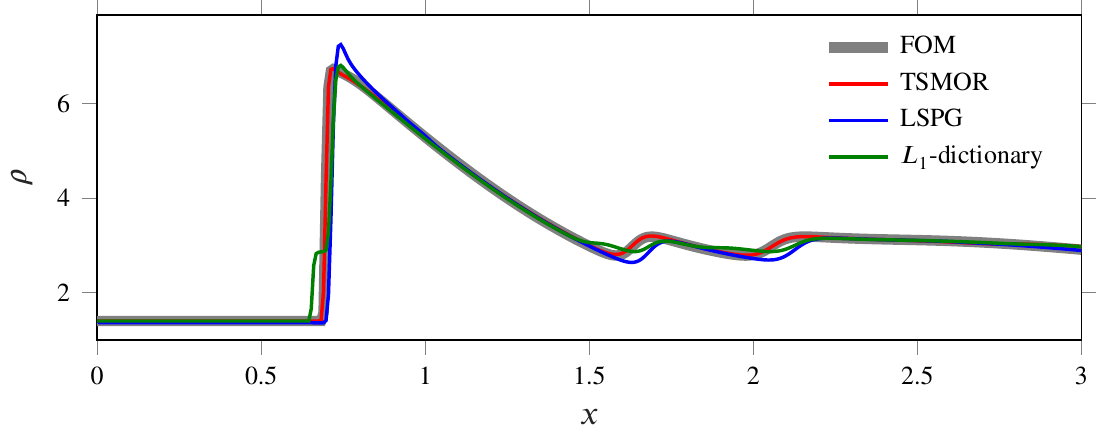}
\caption{$y=0.83$}
\end{subfigure}

\caption{1-D density plots of solutions predicted by FOM, TSMOR, LSPG and $L_1$-dictionary approaches at various $y$-locations}
\label{step1D}
\end{figure}

Next, the performance of the proposed TSMOR approach is compared to several existing MOR techniques. Fig.~\ref{step_pred} illustrates the predictive capabilities of several MOR
approaches for the parameter $\mu^*=3.4$. The FOM density solution is
shown in Fig.~\ref{step_FOM} while the new proposed TSMOR solution using 2 local bases corresponding to two nearby snapshots is shown in
Fig.~\ref{step_TSMOR}. Fig.~\ref{step_LSPG} corresponds to the solution obtained by LSPG using 4 POD modes of the snapshot matrix $\bm{M}$. Fig.~\ref{step_L1} corresponds to $L_1$-dictionary approach using 2 local bases or dictionary elements. To compare these solutions qualitatively, the predicted density distributions at various $y$-locations for the FOM, TSMOR, LSPG and $L_1$-dictionary approaches are shown in
Fig.~\ref{step1D}. It can be observed that the proposed TSMOR approach significantly outperforms
LSPG and $L_1$-dictionary methods.


\begin{figure}[h!]
\centering
\includegraphics[scale=1]{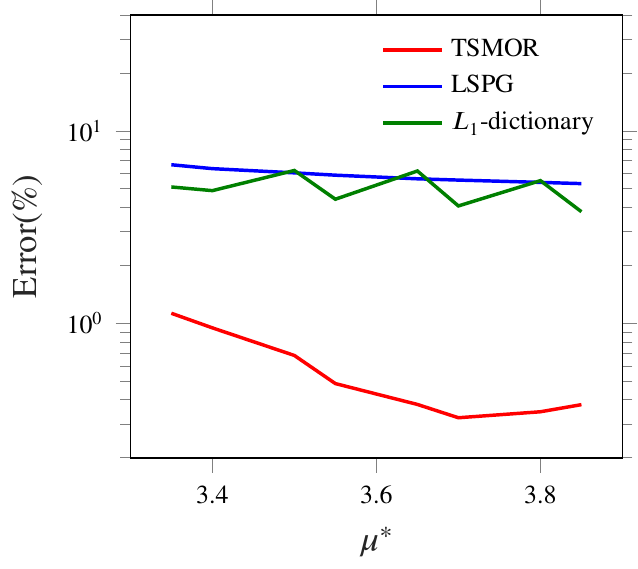}
\caption{Performance comparison between TSMOR, LSPG and $L_1$-dictionary approaches for solution predictions at various parameters}
\label{step_error}
\end{figure}

Finally, the relative solution error between the FOM solution and
predicted solution using TSMOR, LSPG and $L_1$-dictionary approaches across the entire parameter range of interest is
given in Fig.~\ref{step_error}. For
this case, predictions are made at two uniformly distributed parameters in every interval of $\mu_s$. It can be observed that the solutions predicted using TSMOR have an average error of only $0.6\%$ as compared to $5.9\%$ in LSPG and $5.0\%$ in $L_1$-dictionary approach.

\subsection{Jet diffusion flame in a combustor}
\subsubsection{Problem description}

\begin{figure}[h]
\centering
\includegraphics[scale=0.5]{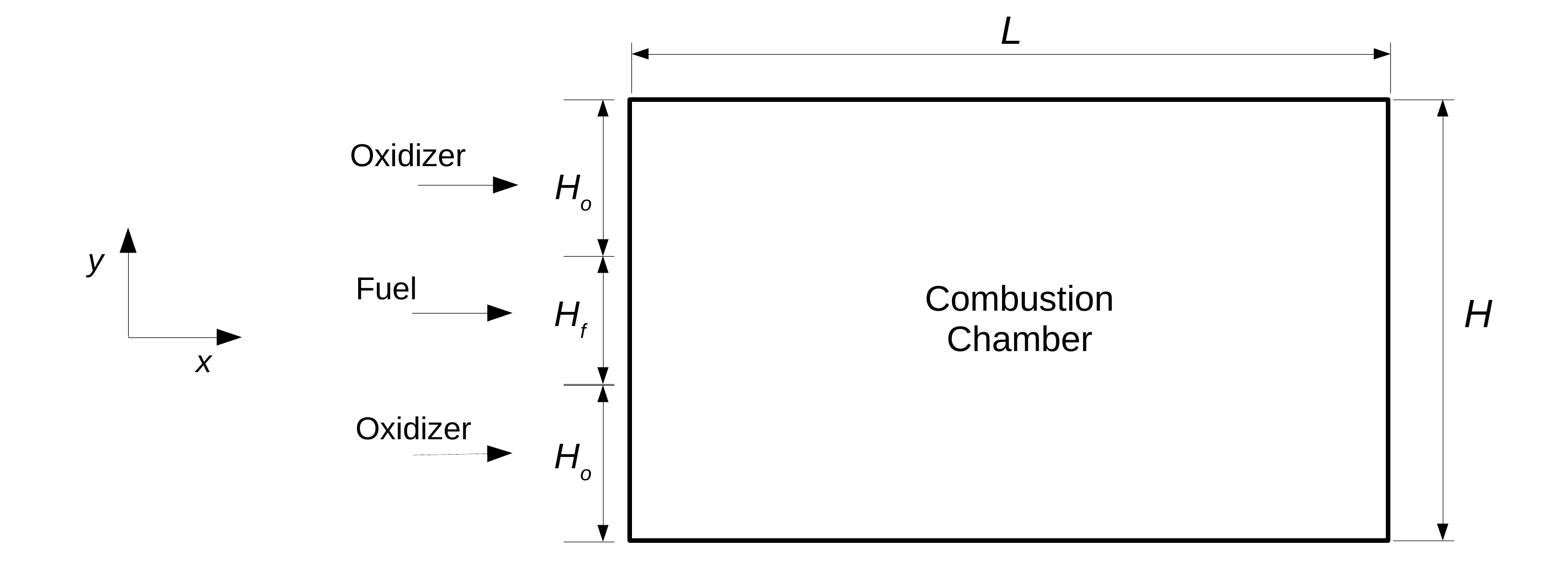}
\caption{Schematic of the combustion chamber}
\label{combustor}
\end{figure}

This problem consists of jets of fuel and oxidizer injected into a combustion chamber as shown in Fig.~\ref{combustor}. Length ($L$) and height ($H$) of the chamber are 18 \si{mm} and 9 \si{mm} respectively. The width of fuel and oxidizer inlets are denoted by $H_f$ and $H_o$ respectively. Inside the chamber, the fuel and oxidizer diffuse to form a diffusion flame where the combustion reaction is governed by an advection-diffusion type governing equation:
\begin{equation}
\nabla\cdot(\bm{W}\bm{w}(\bm{\mu}))-\nabla\cdot(\nu\nabla \bm{w}(\bm{\mu}))+\bm{F}(\bm{w}(\bm{\mu}))=0 \hspace{1cm} x\in[0,L], \hspace{0.3cm} y\in[0,H]
\end{equation}
where the state variable $\bm{w}(\bm{\mu})$ represents the concentration of fuel in the chamber, $\bm{W}=W_x\hat{i} + W_y\hat{j}$ is the velocity field and $\nu$ is the diffusion coefficient. The nonlinear reaction term $\bm{F}(\bm{w}(\bm{\mu}))$ is of Arrhenius type given by:
\begin{equation}
\bm{F}(\bm{w}(\bm{\mu}))=A\bm{w}(\bm{\mu})(c-\bm{w}(\bm{\mu}))e^{-E/(d-\bm{w}(\bm{\mu}))}
\end{equation}
where $c$, $d$, $A$ and $E$ are constants and $(c-\bm{w}(\bm{\mu}))$ represents the oxidizer concentration. Dirichlet boundary condition is prescribed at the inlet:
$$\bm{w}(0,y;\bm{\mu})=0 \hspace{1cm} y\in[0,H_o)$$
$$\bm{w}(0,y;\bm{\mu})=c \hspace{1cm} y\in[H_o,H_o+H_f]$$
$$\bm{w}(0,y;\bm{\mu})=0 \hspace{1cm} y\in(H_o+H_f,H]$$
whereas homogeneous Neumann boundary conditions are prescribed on other boundaries.

\begin{figure}[htbp!]
\centering
\begin{subfigure}[b]{0.45\textwidth}
\centering
\includegraphics[scale=1]{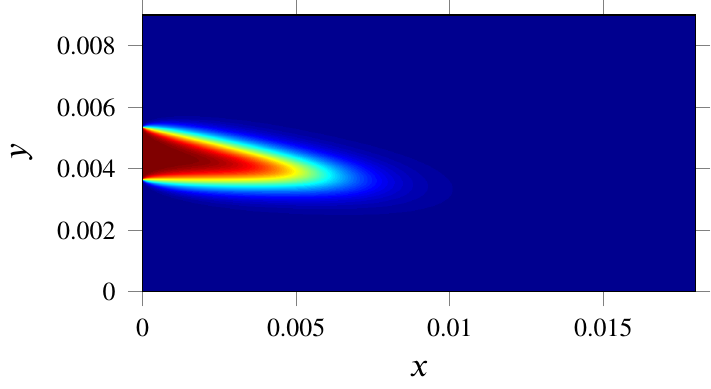}
\vspace{-0.25cm}
\caption{Contour plot at $\bm{\mu}=[-0.02,7,0.467]$}
\label{known11_1}
\end{subfigure}
\begin{subfigure}[b]{0.45\textwidth}
\centering
\includegraphics[scale=1]{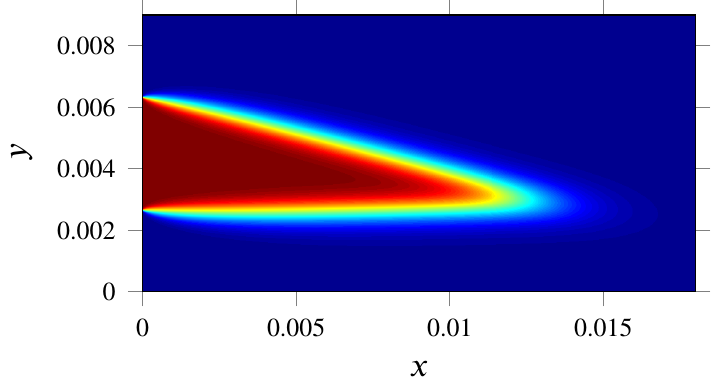}
\vspace{-0.25cm}
\caption{Contour plot at $\bm{\mu}=[-0.02,7,1.364]$}
\label{known12_1}
\end{subfigure}
\begin{subfigure}[b]{0.45\textwidth}
\centering
\includegraphics[scale=1]{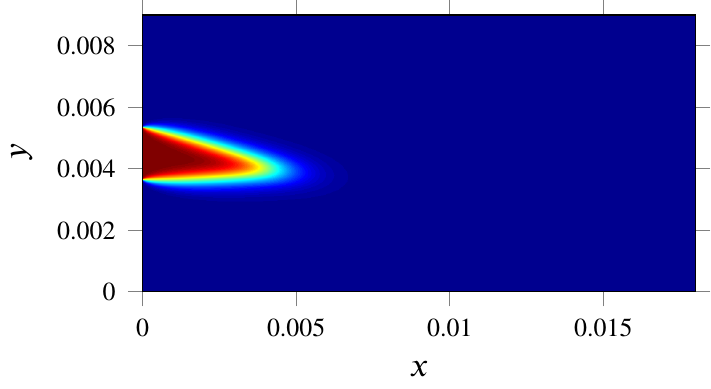}
\vspace{-0.25cm}
\caption{Contour plot at $\bm{\mu}=[-0.02,7.6,0.467]$}
\label{known11_2}
\end{subfigure}
\begin{subfigure}[b]{0.45\textwidth}
\centering
\includegraphics[scale=1]{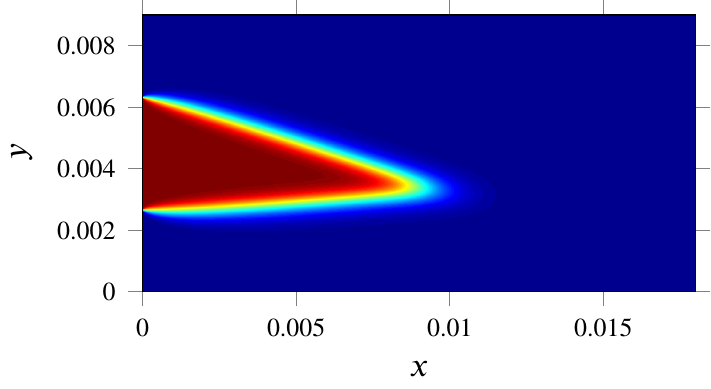}
\vspace{-0.25cm}
\caption{Contour plot at $\bm{\mu}=[-0.02,7.6,1.364]$}
\label{known12_2}
\end{subfigure}
\begin{subfigure}[b]{0.45\textwidth}
\centering
\includegraphics[scale=1]{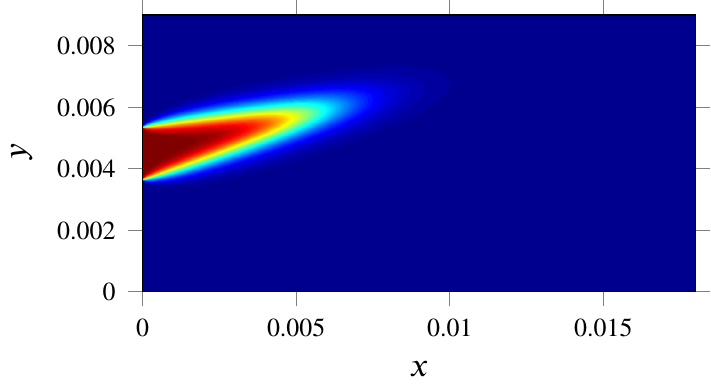}
\vspace{-0.25cm}
\caption{Contour plot at $\bm{\mu}=[0.04,7,0.467]$}
\label{known11_3}
\end{subfigure}
\begin{subfigure}[b]{0.45\textwidth}
\centering
\includegraphics[scale=1]{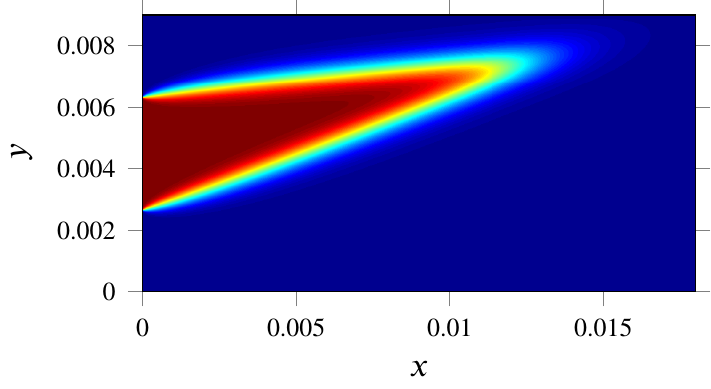}
\vspace{-0.25cm}
\caption{Contour plot at $\bm{\mu}=[0.04,7,1.364]$}
\label{known12_3}
\end{subfigure}
\begin{subfigure}[b]{0.45\textwidth}
\centering
\includegraphics[scale=1]{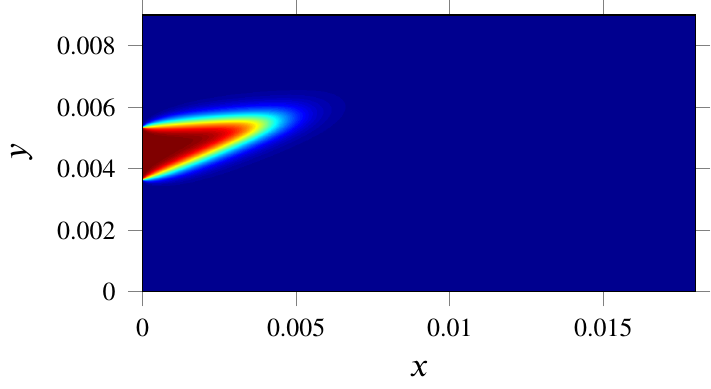}
\vspace{-0.25cm}
\caption{Contour plot at $\bm{\mu}=[0.04,7.6,0.467]$}
\label{known11_4}
\end{subfigure}
\begin{subfigure}[b]{0.45\textwidth}
\centering
\includegraphics[scale=1]{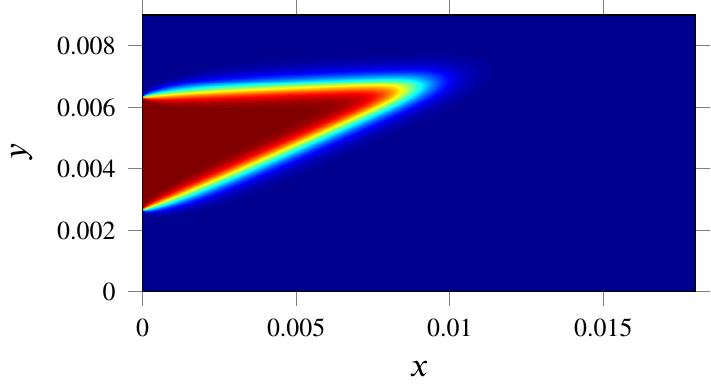}
\vspace{-0.25cm}
\caption{Contour plot at $\bm{\mu}=[0.04,7.6,1.364]$}
\label{known12_4}
\end{subfigure}
\caption{Fuel concentration $\bm{w}(\bm{\mu})$ contours computed at eight different corners of the 3-D parameter space $\mathcal{D}$}
\label{comb8}
\end{figure}

In this problem, we consider parameter variations in three dimensions where $\bm{\mu}=[W_y, ln(A), r]$ consists of the velocity field in $y$-direction $W_y$, Arrhenius parameter $ln(A)$ and ratio of fuel to oxidizer inlet widths $r=H_f/H_o$. $W_y$, $ln(A)$ and $r$ influence the direction, length and width of the flame, respectively. The values of the remaining constants are: ${W}_x=0.17 \si{m/s}$, $\nu=5\times 10^{-6} \si{m^2/s}$, $c=0.2$, $d=0.24$ and $E=0.1091$. The equations are discretized in
space using a second-order, central finite difference scheme on a uniform
Cartesian grid which is divided into 1 million grid points with $\Delta x=\Delta
y=10^{-5} \si{mm}$. A fine discretization is chosen to demonstrate the speedup associated with the hyper-reduction approach. The
resulting equations are solved using Newton's method until convergence of 10 orders of magnitude. Fig.~\ref{comb8} shows the fuel concentration contours computed at eight different corners of the 3-D parameter space. Clear and distinct flame fronts having different directions, lengths and widths can be observed in these plots.

\subsubsection{Implementation of TSMOR}
A snapshot matrix $\bm{M}$ containing 48 snapshots is generated on a $4\times 4 \times 3$ grid in parameter space $\mathcal{D}$ at parameters $W_y\times ln(A) \times r \equiv [-0.02,0,0.02,0.04]\times [7.0,7.2,7.4,7.6] \times [0.467, 0.846, 1.364]$. The coefficients of the polynomial expansion~\eqref{transport_polynomial5} $\bm{c}^{s}$ for each snapshot are computed offline by solving the training error minimization~\eqref{MOR11}. Since the parameter $W_y$ changes the angle of the flame, a strong coupling between $x$ and $y$ coordinates is required. As a result, the basis of the polynomial expansion, $f_{x_p}(x,y)$ and $f_{y_p}(x,y)$, for the $x$ and $y$ transport fields, $c_{s_x}(x,y;\bm{\Delta \mu})$ and $c_{s_y}(x,y;\bm{\Delta \mu})$, respectively, are chosen to be a combination of coupled and decoupled Fourier sine series with $m$ modes. Here we take $f_{x_p}(x,y)=f_{y_p}(x,y)=f_p(x,y)$:
\begin{equation}
\begin{aligned}
&m=1 \quad \bm{f}(x,y)=\sin\left(\frac{\pi {x}}{2L}\right)\\
&m=2 \quad \bm{f}(x,y)=\left[\bm{f}(x,y)\bigg\rvert_{m=1}, \sin\left(\frac{\pi {x}}{L}\right), \sin\left(\frac{\pi {y}}{H}\right) \right]\\
&m=3 \quad \bm{f}(x,y)=\left[\bm{f}(x,y)\bigg\rvert_{m=2}, \sin\left(\frac{2\pi {x}}{L}\right), \sin\left(\frac{\pi {x}}{L}\right)\sin\left(\frac{\pi {y}}{H}\right), \sin\left(\frac{2\pi {y}}{H}\right) \right]\\
&m=4 \quad \bm{f}(x,y)=\left[\bm{f}(x,y)\bigg\rvert_{m=3}, \sin\left(\frac{3\pi {x}}{L}\right), \sin\left(\frac{2\pi {x}}{L}\right)\sin\left(\frac{\pi {y}}{H}\right), \sin\left(\frac{\pi {x}}{L}\right)\sin\left(\frac{2\pi {y}}{H}\right), \sin\left(\frac{3\pi {y}}{H}\right) \right]
\end{aligned}
\label{comb_transport}
\end{equation}
Note that the basis functions in Eq.~\eqref{comb_transport} is cumulative in the sense that for $m>1$, the basis includes the basis corresponding to the previous modes as well. 
Since the inlet boundary condition for this problem is discontinuous, boundary conditions on the transported snapshots are enforced by enforcing appropriate conditions on the transports, as described in remark 2 of \S~\ref{Implementation}.
More specifically, to avoid extrapolation in the $x$-direction, the following condition is imposed on the $x$-transports at the inlet: $\bm{c}_{s_x}(x=0,y;\bm{\Delta \mu})=0$. This condition is enforced by setting the coefficients of decoupled basis $\sin(i \pi y/H)=0$ for $i=1,2,3$ in the expansion of $\bm{c}_{s_x}(x,y;\bm{\Delta \mu})$. Note that this condition is not enforced on $y$-transports $\bm{c}_{s_y}(x,y;\bm{\Delta \mu})$.
For this multi-dimensional parameter problem, $g_q(\bm{\Delta \mu})$  is given by:
\begin{equation}
\bm{g}(\bm{\Delta \mu})=\left[\Delta \mu_{1}, \hspace{0.2cm} \Delta \mu_{2}, \hspace{0.2cm} \Delta \mu_{3}, \hspace{0.2cm} \Delta \mu_{1}\Delta \mu_{2}, \hspace{0.2cm} \Delta \mu_{2}\Delta \mu_{3}, \hspace{0.2cm} \Delta \mu_{3}\Delta \mu_{1}, \hspace{0.2cm} \Delta \mu_{1}\Delta \mu_{2}\Delta \mu_{3} \right]
\end{equation}

\begin{figure}[h]
\centering
\includegraphics[scale=1]{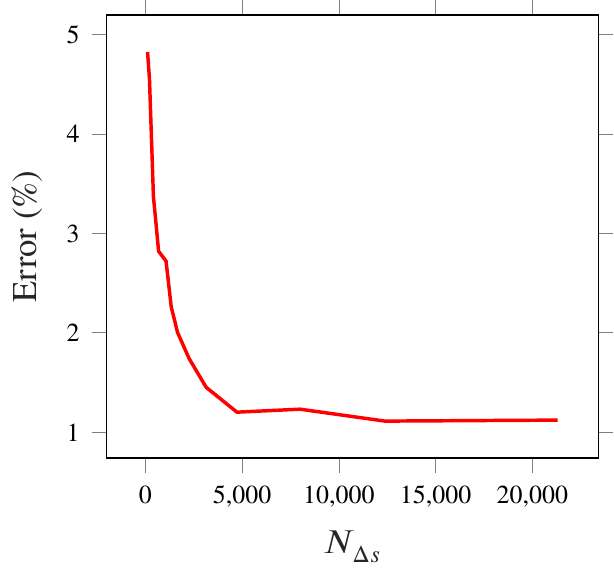}
\caption{Grid convergence study for the offline stage of the proposed TSMOR approach}
\label{comb_grid_conv}
\end{figure}

Transformation of the structured Cartesian grid based on the coupled
transports~\eqref{comb_transport} leads to a non-tensor grid. Hence,
interpolation from the non-tensor transported grid to the original grid
must be performed using the \verb=scatteredInterpolant= algorithm in Matlab. 
Unfortunately, this interpolation scheme is computationally prohibitive for large
grids. Therefore, during the offline stage, the snapshots are uniformly
downsampled by a factor $\Delta s$ in $x$ and $y$ coordinates leading to a coarser grid with $N_{\Delta s}$ points. A grid convergence study is conducted to identify $\Delta s$ and $N_{\Delta s}$.
Fig.~\ref{comb_grid_conv} plots the relative error in the TSMOR solutions predicted at $\mu^*=[-0.01,7.1,0.644]$ for increasing $N_{\Delta s}$. It can be observed that the reduction in error is minimal after 3000 grid points which corresponds to a downsampling factor of 25. Hence, the snapshots are uniformly
downsampled by a factor of 25. Note that this downsampling process is conducted \emph{only} during the offline stage.

\begin{figure}[h!]
\centering
\begin{subfigure}[t]{0.4\textwidth}
\centering
\includegraphics[scale=1]{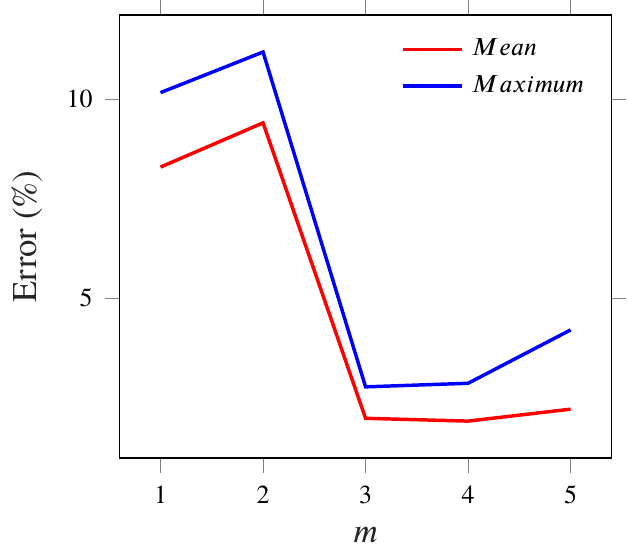}
\caption{Plot of relative error v/s number of Fourier modes $m$ for TSMOR predicted solutions}
\label{comb_Fourier_conv1}
\end{subfigure}\hspace{1cm}
\begin{subfigure}[t]{0.4\textwidth}
\centering
\includegraphics[scale=1]{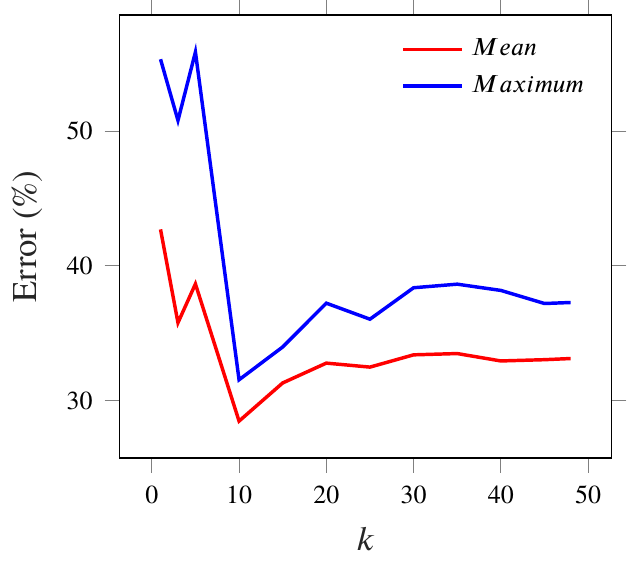}
\caption{Plot of relative error v/s number of POD basis $k$  for LSPG predicted solutions}
\label{comb_Fourier_conv2}
\end{subfigure}
\caption{Convergence plot of relative error for TSMOR and LSPG predicted solutions}
\label{comb_Fourier_conv}
\end{figure}

Next, convergence of the proposed TSMOR approach with respect to the number of Fourier modes $m$ is studied by predicting new solutions in the predictive regime $\mu^*$. Fig.~\ref{comb_Fourier_conv1} shows the mean and maximum relative error in the TSMOR solutions predicted at 18 uniformly distributed parameters in the parameter space $\mathcal{D}$ for different numbers of Fourier modes $m$. It can be observed that as the number of Fourier modes increases, the error converges to a low value of 1.91\%. The TSMOR convergence plot is compared to the convergence of LSPG approach with respect to the number of POD basis $k$ of the snapshot matrix $\bm{M}$. Similar to Fig.~\ref{comb_Fourier_conv1}, Fig.~\ref{comb_Fourier_conv2} displays the relative errors in the LSPG solutions predicted at the same set of parameters for different numbers of bases. It can be observed that, the error converges only to 30\% even though all 48 POD basis were used for prediction. Hereafter, all the TSMOR predicted results presented for this problem are produced with 4 Fourier modes.

Next, the TSMOR approach is equipped with the hyper-reduction strategy mentioned in \S~\ref{HR}. First, 70 collocation points are obtained by employing the DEIM algorithm. Second, these points are augmented with 30 uniformly distributed inlet points. Finally, an additional $\hat{n}_w\approx n_w \times 4=400$ points $\hat{\varepsilon}$ are included to enable the evaluation of the residuals via the central finite difference scheme.

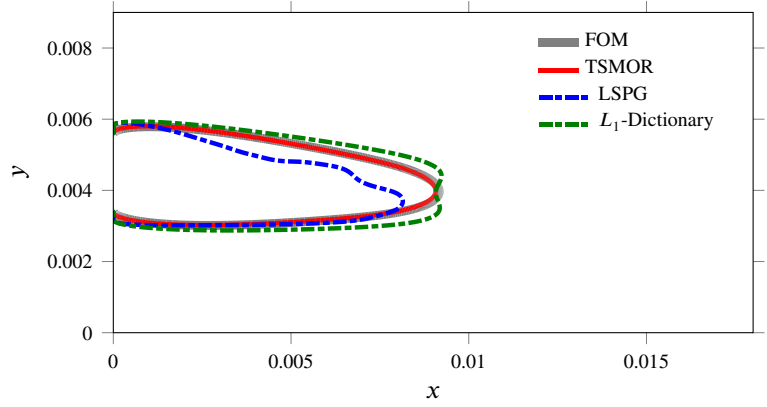
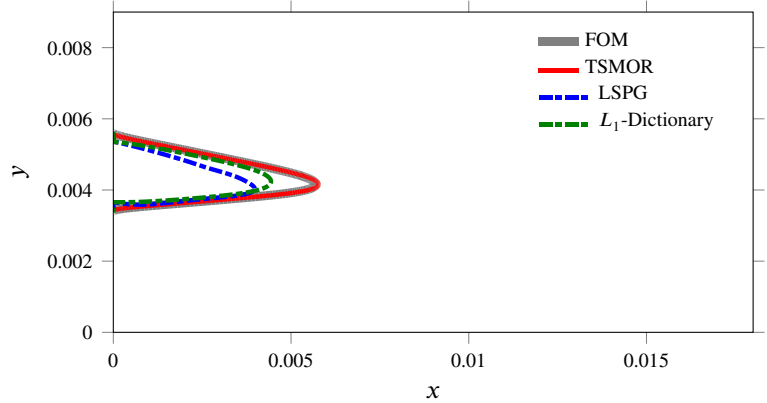
\begin{figure}
\centering
\begin{subfigure}[b]{1\textwidth}
\centering
\includegraphics[scale=1]{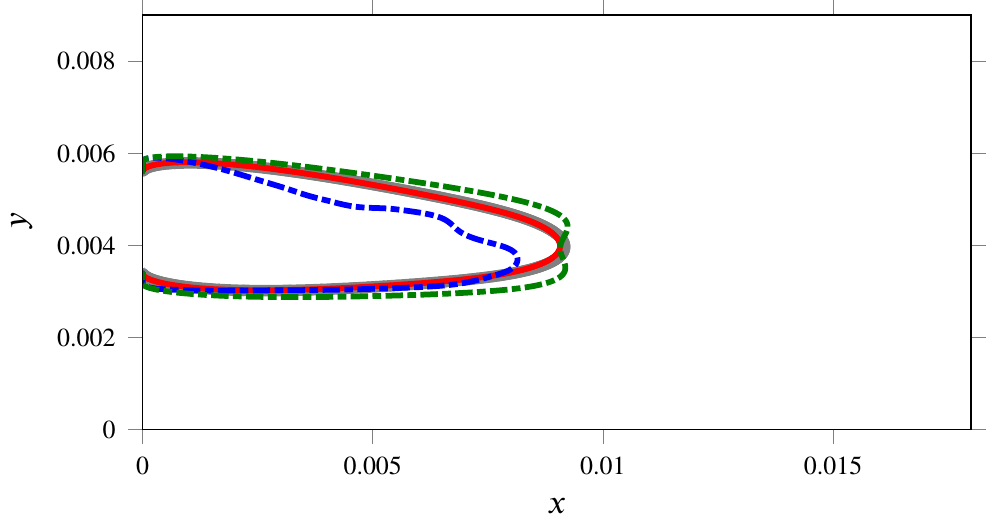}
\put(-85,135){\color{gray}\rule{0.03\textwidth}{3pt} \color{black} \footnotesize FOM}
\put(-85,125){\color{red}\rule{0.03\textwidth}{1.5pt} \color{black} \footnotesize TSMOR}
\put(-85,115){\color{blue}\hdashrule[0ex][x]{0.04\textwidth}{1.5pt}{5pt 1pt 2.3pt 1pt} \color{black} \footnotesize LSPG}
\put(-85,105){\color{OliveGreen}\hdashrule[0ex][x]{0.04\textwidth}{1.5pt}{5pt 1pt 2.3pt 1pt} \color{black} \footnotesize $L_1$-Dictionary}
\caption{Contour plot for $\bm{w}(\bm{\mu}^*)=0.018$}
\vspace{0.25cm}
\label{known11_5}
\end{subfigure}
\begin{subfigure}[b]{1\textwidth}
\centering
\includegraphics[scale=1]{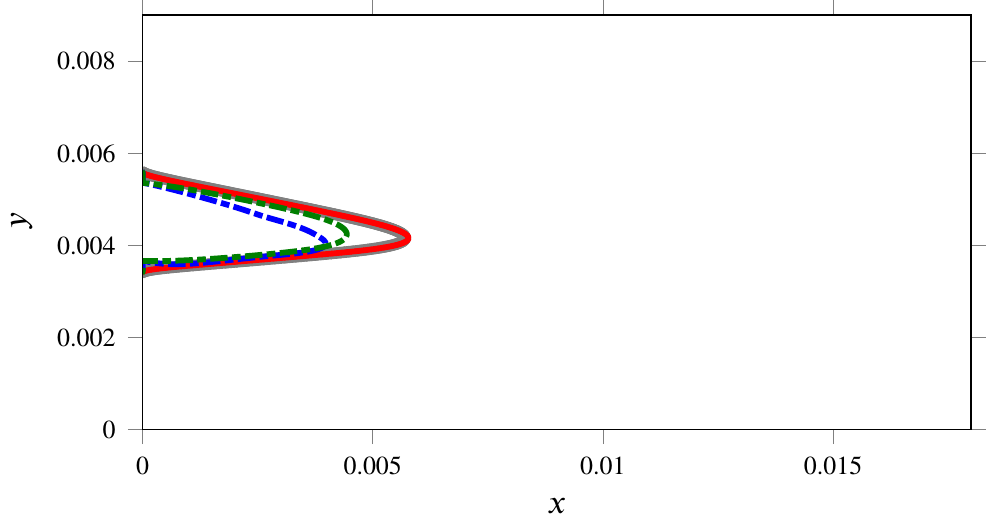}
\put(-85,135){\color{gray}\rule{0.03\textwidth}{3pt} \color{black} \footnotesize FOM}
\put(-85,125){\color{red}\rule{0.03\textwidth}{1.5pt} \color{black} \footnotesize TSMOR}
\put(-85,115){\color{blue}\hdashrule[0ex][x]{0.04\textwidth}{1.5pt}{5pt 1pt 2.3pt 1pt} \color{black} \footnotesize LSPG}
\put(-85,105){\color{OliveGreen}\hdashrule[0ex][x]{0.04\textwidth}{1.5pt}{5pt 1pt 2.3pt 1pt} \color{black} \footnotesize $L_1$-Dictionary}
\caption{Contour plot for $\bm{w}(\bm{\mu}^*)=0.15$}
\label{known12_5}
\end{subfigure}
\caption{Case 1: Comparison of predicted solutions at $\bm{\mu^*}=[-0.01, 7.1, 0.643]$ using TSMOR, LSPG and $L_1$-Dictionary approach with FOM}
\label{known1_comb}
\end{figure}

\begin{figure}[h]
\centering
\begin{subfigure}[b]{1\textwidth}
\centering
\includegraphics[scale=1]{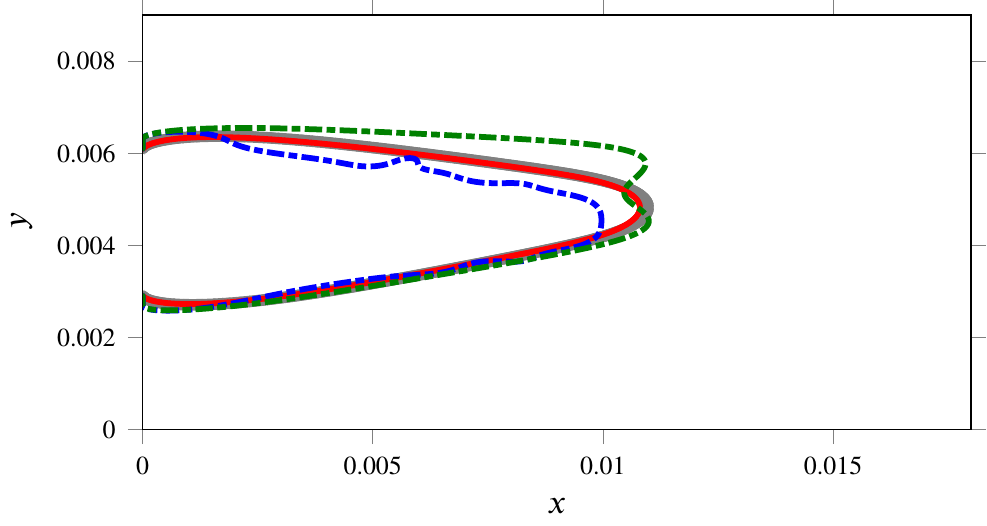}
\put(-85,135){\color{gray}\rule{0.03\textwidth}{3pt} \color{black} \footnotesize FOM}
\put(-85,125){\color{red}\rule{0.03\textwidth}{1.5pt} \color{black} \footnotesize TSMOR}
\put(-85,115){\color{blue}\hdashrule[0ex][x]{0.04\textwidth}{1.5pt}{5pt 1pt 2.3pt 1pt} \color{black} \footnotesize LSPG}
\put(-85,105){\color{OliveGreen}\hdashrule[0ex][x]{0.04\textwidth}{1.5pt}{5pt 1pt 2.3pt 1pt} \color{black} \footnotesize $L_1$-Dictionary}
\caption{Contour plot for $\bm{w}(\bm{\mu}^*)=0.018$}
\vspace{0.25cm}
\label{known21}
\end{subfigure}
\begin{subfigure}[b]{1\textwidth}
\centering
\includegraphics[scale=1]{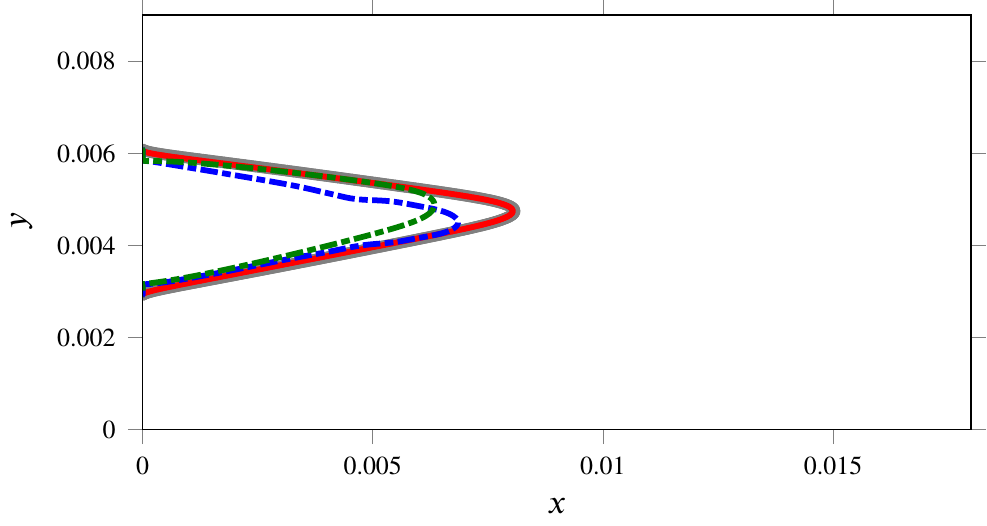}
\put(-85,135){\color{gray}\rule{0.03\textwidth}{3pt} \color{black} \footnotesize FOM}
\put(-85,125){\color{red}\rule{0.03\textwidth}{1.5pt} \color{black} \footnotesize TSMOR}
\put(-85,115){\color{blue}\hdashrule[0ex][x]{0.04\textwidth}{1.5pt}{5pt 1pt 2.3pt 1pt} \color{black} \footnotesize LSPG}
\put(-85,105){\color{OliveGreen}\hdashrule[0ex][x]{0.04\textwidth}{1.5pt}{5pt 1pt 2.3pt 1pt} \color{black} \footnotesize $L_1$-Dictionary}
\caption{Contour plot for $\bm{w}(\bm{\mu}^*)=0.15$}
\label{known22}
\end{subfigure}
\caption{Case 2: Comparison of predicted solutions at $\bm{\mu^*}=[0.005, 7.3, 1.083]$ using TSMOR, LSPG and $L_1$-Dictionary approach with FOM}
\label{known23}
\end{figure}

\begin{figure}[h]
\centering
\begin{subfigure}[b]{1\textwidth}
\centering
\includegraphics[scale=1]{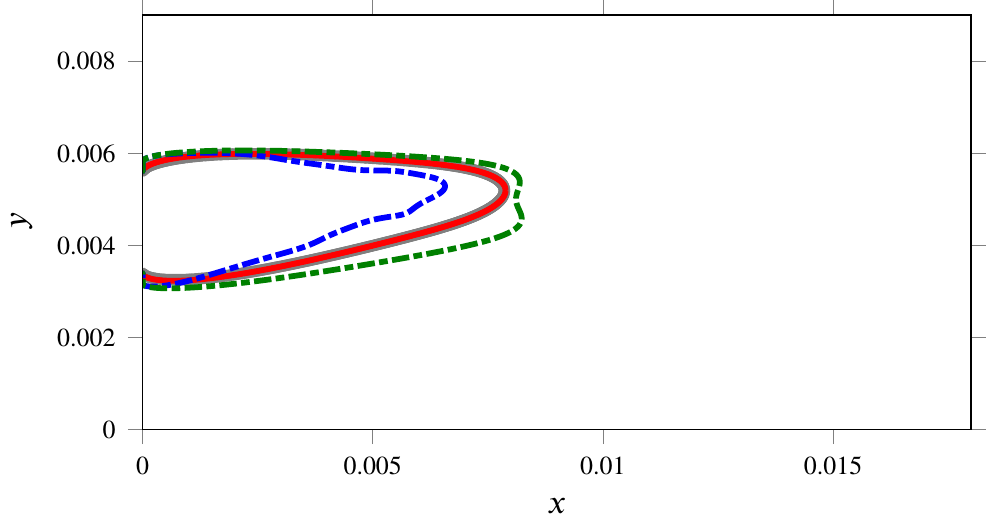}
\put(-85,135){\color{gray}\rule{0.03\textwidth}{3pt} \color{black} \footnotesize FOM}
\put(-85,125){\color{red}\rule{0.03\textwidth}{1.5pt} \color{black} \footnotesize TSMOR}
\put(-85,115){\color{blue}\hdashrule[0ex][x]{0.04\textwidth}{1.5pt}{5pt 1pt 2.3pt 1pt} \color{black} \footnotesize LSPG}
\put(-85,105){\color{OliveGreen}\hdashrule[0ex][x]{0.04\textwidth}{1.5pt}{5pt 1pt 2.3pt 1pt} \color{black} \footnotesize $L_1$-Dictionary}
\caption{Contour plot for $\bm{w}(\bm{\mu}^*)=0.018$}
\vspace{0.25cm}
\label{known31}
\end{subfigure}
\begin{subfigure}[b]{1\textwidth}
\centering
\includegraphics[scale=1]{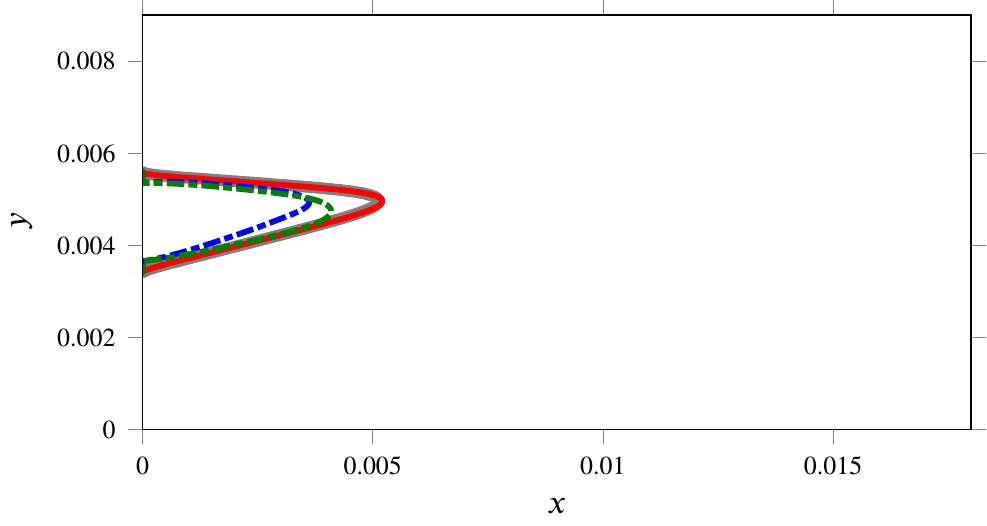}
\put(-85,135){\color{gray}\rule{0.03\textwidth}{3pt} \color{black} \footnotesize FOM}
\put(-85,125){\color{red}\rule{0.03\textwidth}{1.5pt} \color{black} \footnotesize TSMOR}
\put(-85,115){\color{blue}\hdashrule[0ex][x]{0.04\textwidth}{1.5pt}{5pt 1pt 2.3pt 1pt} \color{black} \footnotesize LSPG}
\put(-85,105){\color{OliveGreen}\hdashrule[0ex][x]{0.04\textwidth}{1.5pt}{5pt 1pt 2.3pt 1pt} \color{black} \footnotesize $L_1$-Dictionary}
\caption{Contour plot for $\bm{w}(\bm{\mu}^*)=0.15$}
\label{known32}
\end{subfigure}
\caption{Case 3: Comparison of predicted solutions at $\bm{\mu^*}=[0.015, 7.35, 0.643]$ using TSMOR, LSPG and $L_1$-Dictionary approach with FOM}
\label{known3}
\end{figure}

\begin{figure}[h]
\centering
\begin{subfigure}[b]{1\textwidth}
\centering
\includegraphics[scale=1]{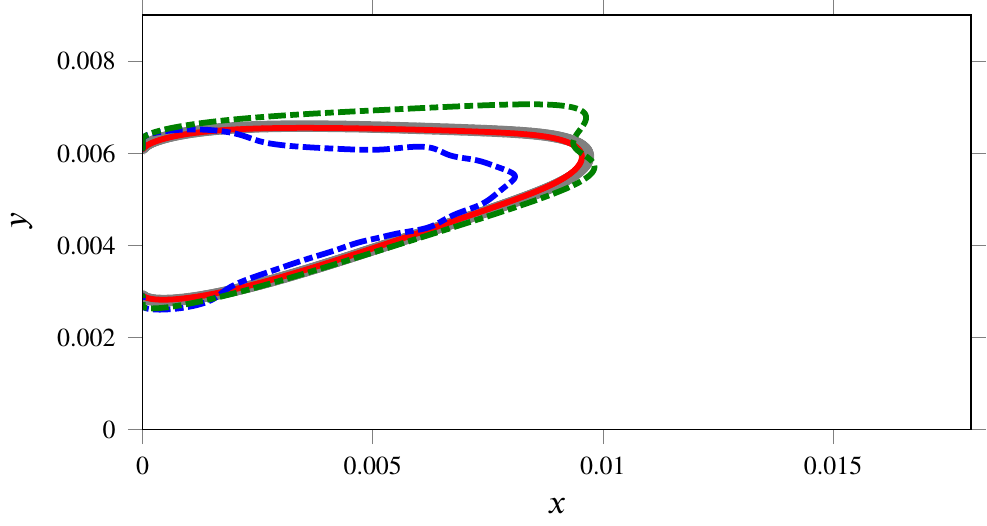}
\put(-85,135){\color{gray}\rule{0.03\textwidth}{3pt} \color{black} \footnotesize FOM}
\put(-85,125){\color{red}\rule{0.03\textwidth}{1.5pt} \color{black} \footnotesize TSMOR}
\put(-85,115){\color{blue}\hdashrule[0ex][x]{0.04\textwidth}{1.5pt}{5pt 1pt 2.3pt 1pt} \color{black} \footnotesize LSPG}
\put(-85,105){\color{OliveGreen}\hdashrule[0ex][x]{0.04\textwidth}{1.5pt}{5pt 1pt 2.3pt 1pt} \color{black} \footnotesize $L_1$-Dictionary}
\caption{Contour plot for $\bm{w}(\bm{\mu}^*)=0.018$}
\vspace{0.25cm}
\label{known41}
\end{subfigure}
\begin{subfigure}[b]{1\textwidth}
\centering
\includegraphics[scale=1]{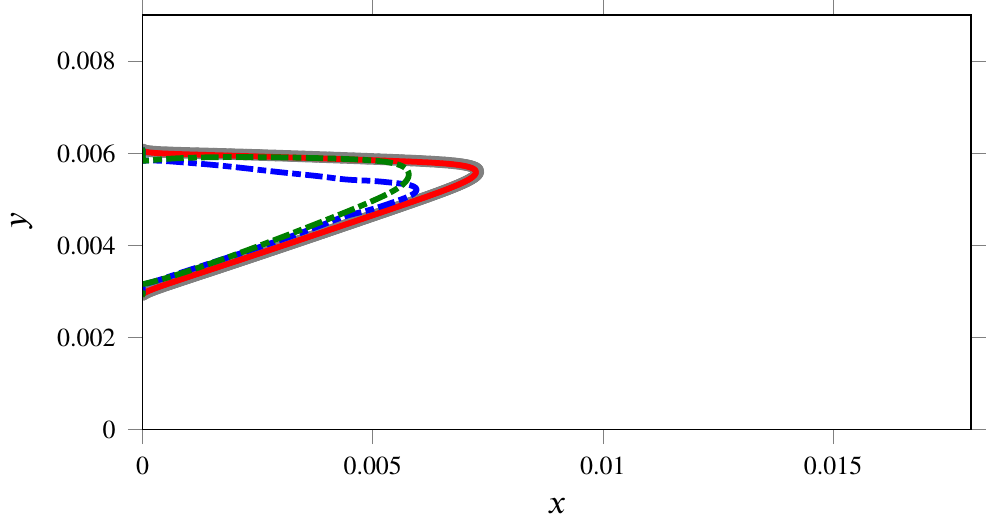}
\put(-85,135){\color{gray}\rule{0.03\textwidth}{3pt} \color{black} \footnotesize FOM}
\put(-85,125){\color{red}\rule{0.03\textwidth}{1.5pt} \color{black} \footnotesize TSMOR}
\put(-85,115){\color{blue}\hdashrule[0ex][x]{0.04\textwidth}{1.5pt}{5pt 1pt 2.3pt 1pt} \color{black} \footnotesize LSPG}
\put(-85,105){\color{OliveGreen}\hdashrule[0ex][x]{0.04\textwidth}{1.5pt}{5pt 1pt 2.3pt 1pt} \color{black} \footnotesize $L_1$-Dictionary}
\caption{Contour plot for $\bm{w}(\bm{\mu}^*)=0.15$}
\label{known42}
\end{subfigure}
\caption{Case 4: Comparison of predicted solutions at $\bm{\mu^*}=[0.025, 7.5, 1.083]$ using TSMOR, LSPG and $L_1$-Dictionary approach with FOM}
\label{known4}
\end{figure}

To demonstrate the performance of the proposed TSMOR approach, solutions are predicted at four different predictive regimes in the parameter space $\mathcal{D}$ and compared with FOM and several other MOR techniques. Parameters corresponding to these test cases are tabulated in Table.~\ref{cases}. Figs.~\ref{known1_comb}-~\ref{known4} illustrate the predictive capabilities of several MOR approaches for these test cases. Contour plots at contour levels $\bm{w}(\bm{\mu}^*)=0.018$ and $\bm{w}(\bm{\mu}^*)=0.15$ are displayed in these figures. The FOM is given by the gray lines while the new proposed hyper-reduced TSMOR approach using 8 local basis is given by the red lines. The blue lines correspond to the solution obtained by LSPG using 48 POD modes of the snapshot matrix $\bm{M}$. The green lines correspond to $L_1$-dictionary approach using 8 local basis or dictionary elements. The proposed TSMOR approach predicts a solution which accurately matches the FOM solutions in all the four cases. In contrast, both LSPG and $L_1$-dictionary methods fail to capture the flame-front. 

\begin{table}[h]
\begin{center}
\begin{tabular}{ |>{\centering\arraybackslash}m{2.5cm}|>{\centering\arraybackslash}m{2.5cm}|>{\centering\arraybackslash}m{2.5cm}|>{\centering\arraybackslash}m{2.5cm}|}
 \hline 
 $\bm{\mu}^*$ & $\bm{\mu}_1^*\equiv W_y$ & $\bm{\mu}_2^* \equiv ln(A)$ & $\bm{\mu}_3^* \equiv r$ \\
\hline
 Case 1 & -0.01 & 7.1 & 0.643 \\ \hline
 Case 2 & 0.005 & 7.3 & 1.083 \\ \hline
 Case 3 & 0.015 & 7.35 & 0.643 \\ \hline
 Case 4 & 0.025 & 7.5 & 1.083 \\ 
 \hline
\end{tabular}
\end{center}
\caption{Table of four predictive test cases}
\label{cases}
\end{table}

The ROM errors for the various MOR approaches are summarized in
Table~\ref{table_error}. Solutions predicted using TSMOR
have an average error of only 1.92\% as compared to 29.02\% in LSPG and 17.53\% in
the $L_1$-dictionary approach. 
The table also provides the error in the POD-projected solution which is
obtained by projecting the FOM onto 48 POD basis of the snapshot matrix $\bm{M}$.
The average error for the POD-projected solution is 8.65\%. Thus, TSMOR significantly
outperforms LSPG and $L_1$-dictionary approaches and it is 3-4 times better than
POD-projected solutions.

\begin{table}[h!]
\begin{center}
\begin{tabular}{ |>{\centering\arraybackslash}m{2.5cm}|>{\centering\arraybackslash}m{2.5cm}|>{\centering\arraybackslash}m{2.5cm}|>{\centering\arraybackslash}m{2.5cm}|>{\centering\arraybackslash}m{2.5cm}| }
 \hline 
 Model & TSMOR+HR \newline (8 local basis) & POD-Projection \newline (48 global basis) & LSPG \newline (48 global basis)& $L_1$ \newline Dictionary \newline (8 local basis)\\
\hline
 Case 1 & 1.07 & 9.12 & 27.34 & 15.49 \\ \hline
 Case 2 & 2.47 & 8.43 & 23.95 & 18.54 \\ \hline
 Case 3 & 1.24 & 8.72 & 32.74 & 18.25 \\ \hline
 Case 4 & 2.92 & 8.32 & 32.07 & 17.83 \\ 
 \hline
\end{tabular}
\end{center}
\caption{Comparison of relative error (\%) in the predicted solution at $\bm{\mu}^*$ for four different cases}
\label{table_error}
\end{table}
\vspace{-0.5cm}

Finally, wall-times and speed-ups for the FOM and the online stage of hyper-reduced
TSMOR are given in Table.~\ref{comb_walltime}. Here, speed-up is defined as the ratio of wall-times of FOM to the online stage of hyper-reduced TSMOR. TSMOR+HR delivers a speed-up of
two orders of magnitude.

\begin{table}[h!]
\begin{center}
\begin{tabular}{ |>{\centering\arraybackslash}m{2.5cm}|>{\centering\arraybackslash}m{2.5cm}|>{\centering\arraybackslash}m{2.5cm}|>{\centering\arraybackslash}m{2.5cm}|}
 \hline 
 Model & FOM & TSMOR+HR & Speed-up\\
\hline
 Case 1 & 169.01 s & 0.43 s & 393 \\ \hline
 Case 2 & 202.37 s & 0.46 s & 440 \\ \hline
 Case 3 & 166.05 s & 0.63 s & 263 \\ \hline
 Case 4 & 208.23 s & 0.71 s & 293 \\ 
 \hline
\end{tabular}
\end{center}
\caption{CPU wall-times associated with the FOM and online stage of hyper-reduced TSMOR for the prediction at $\bm{\mu}^*$ for four different cases}
\label{comb_walltime}
\end{table}

\section{Limitations of TSMOR}
\label{Limitations}

TSMOR can be expected to provide significant improvements in accuracy over traditional projection-based ROMs for a large class of problems characterized by shocks and sharp gradients whose spatial locations and orientations are strongly parameter dependent. There are, however, a large class of important and very challenging problems where TSMOR -- and indeed most other related methodologies in the literature -- can be expected to fail.

For example, similar to most other model reduction approaches for parameter variations in the literature, TSMOR requires that the topology of the underlying solution varies smoothly with respect to the parameters. In other words, TSMOR cannot be expected to provide accurate predictions of bifurcations. In the case when the changes in solutions with respect to parameter variations are sharp yet still smooth and continuous, an adaptive parameter sampling~\cite{paul2015adaptive} strategy during the offline stage can be expected to mitigate this issue.

Finally, since both the online and offline stages of TSMOR require solutions of non-convex optimization problems (Eq.~\eqref{MOR10} and~\eqref{MOR11}, respectively), convergence to local, sub-optimal minima can be a real problem. However, in principle, this issue can usually be avoided by implementing a global optimization scheme, or -- more typically -- refining and/or adaptively sampling the parameter space during the offline stage.

\section{Conclusions}
\label{Conclusions}

In this manuscript, a transported snapshot model order reduction (TSMOR) method
for predicting new parametric steady state solutions containing moving shocks
and discontinuities is presented. In this method, the solution is approximated by a linear
combination of spatially transported snapshots. The transports are assumed to be smooth in parameter as well as physical space, and hence approximated as a low-order polynomial expansion. The coefficients of the polynomial expansion are obtained by solving a training error minimization problem in the offline stage. The generalized coordinates are derived by solving a residual minimization problem in the online stage. TSMOR is also integrated with hyper-reduction methods to reduce the computational complexity of evaluating the nonlinear residual and parameter dependent basis in the online stage. 

Numerical experiments consist of a 1-D converging-diverging
nozzle problem with throat area as the parameter, a supersonic flow over a
forward facing step with inlet Mach number as the parameter and a multi-dimensional parametric combustion problem with three parameters influencing the length, direction and width of the diffusion flame. For all parameters
considered, TSMOR is demonstrated to significantly outperform traditional
approaches such as those based on linear compression
schemes e.g. LSPG and more recent local basis approaches such as $L_1$-dictionary. Furthermore, speed-up of two and four orders of magnitude for the combustion and nozzle problems are achieved, respectively.

\section{Acknowledgement}
This material is based upon the work supported by the Air Force Office of Scientific Research
under Grant No. FA9550-17-1-0203. 

\bibliographystyle{WileyNJD-AMA}
\bibliography{references.bib}

\end{document}